\documentstyle[12pt]{article}
\newcommand{\be}{\begin{equation}}
\newcommand{\ee}{\end{equation}}
\newcommand{\bea}{\begin{eqnarray}}
\newcommand{\eea}{\end{eqnarray}}
\newcommand{\ol}{\overline}
\newcommand{\ul}{\underline}
\newcommand{\al}{\alpha}
\newcommand{\th}{\theta}
\newcommand{\Th}{\Theta}
\newcommand{\la}{\lambda}
\newcommand{\ga}{\gamma}
\newcommand{\Ga}{\Gamma}
\newcommand{\si}{\sigma}
\newcommand{\rb}{\right]}
\newcommand{\lb}{\left[}
\newcommand{\cR}{\check{R}}
\newcommand{\pol}{\frac{i\pi}{3\la}}
\newcommand{\hsv}{\hspace{7pt}}
\newcommand{\hfv}{\hspace{5pt}}
\newcommand{\fns}{\footnotesize}
\newcommand{\scs}{\scriptsize}
\newcommand{\shs}{\shortstack}
\newcommand{\nn}{\nonumber}
\advance\textheight by \topskip
\begin{document}

\newpage
\begin{titlepage}
\begin{flushright}
DAMTP-95-04\\
hep-th/9501136\\
revised May 95
\end{flushright}
\vspace{1cm}
\begin{center}
{\bf {\LARGE Exact S-Matrices for Bound States\\ of $a_2^{(1)}$ Affine
Toda Solitons}}\\
\vspace{1.5cm}
{\large Georg M. Gandenberger}\footnote{e-mail:
G.M.Gandenberger@damtp.cam.ac.uk} \\
\vspace{3mm}
{\em Department of Applied Mathematics and Theoretical Physics,}\\
{\em Cambridge University}\\
{\em Silver Street, Cambridge, CB3 9EW, U.K.}\\
\vspace{2cm} {\bf{ABSTRACT}}
\end{center}
\begin{quote}
Using Hollowood's conjecture for the S-matrix for
elementary solitons in complex $a_n^{(1)}$ affine Toda field theories
we examine the interactions of bound states of solitons in $a_2^{(1)}$
theory. The elementary solitons can form two different kinds of bound
states: scalar bound states (the so-called breathers), and excited
solitons, which are bound states with non-zero topological charge. We
give explicit expressions of all S-matrix elements involving the
scattering of breathers and excited solitons and examine their
pole structure in detail. It is shown how the poles can
be explained in terms of on-shell diagrams, several of which involve a
generalized Coleman-Thun mechanism.
\end{quote}
\vspace{1cm}
\begin{center}
{\em To appear in:} NUCLEAR PHYSICS B
\end{center}
\vfill

\end{titlepage}

\section{Introduction}

This paper deals with a certain class of so-called two-dimensional
completely integrable massive models, which are relativistic quantum
field theories defined in 1+1 dimensional Minkowski space.
Two-dimensional models were first studied during the seventies and
have attracted increased attention during the last fifteen years.
Zamolodchikov \cite{zamol2}
was the first to notice that integrable theories can be obtained as
special deformations of conformal field theories, where due to the
deformation the conformality is destroyed, but the integrability of
the theory can be preserved and the particles acquire masses.  From
the viewpoint of integrable theories, conformal field
theories can be regarded as limits of certain massive integrable
models and appear as their particular massless limits.

In order to obtain all the on-shell information of a particular
quantum field theory we seek to find an explicit expression for the
S-matrix. However, finding exact solutions and
understanding non-perturbative effects in quantum field theories in
higher dimensions is an extremely difficult task and is
regarded as one of the outstanding problems of theoretical physics.
In quantum field theories defined in 1+1 dimensions, however, this
problem appears to be more manageable. In most cases it is possible
to give an explicit expression for the S-matrix, which is both
self-consistent and consistent with a Lagrangian formulation of the
theory.
The main tool for finding exact S-matrices is the exact S-matrix
program, which was developed in the sixties in order to describe the
strong interaction (see \cite{eden}). In the case of integrable
theories the axioms of exact S-matrix
theory, which will be discussed in section 2, enable us in many cases
to find explicit expressions for the S-matrices up to a so-called
CDD-factor.

In this paper we consider a special class of integrable
two-dimensional theories, the so-called affine Toda field theories
(ATFTs).
Affine Toda field theories are a family of integrable quantum
field theories in 1+1 dimensional Minkowski space. They are
defined by their Lagrangian density:
\be
{\cal L} =
\frac12(\partial_{\mu}\phi^a)(\partial^{\mu}\phi^a) -
\frac{m^2}{\tilde{\beta}^2}
\sum_{j=0}^r n_je^{\tilde{\beta}\al_j\phi}. \label{lagrangian}
\ee
where $\phi = (\phi^1,...,\phi^r)$ is a r-dimensional scalar field,
the $\al_i$ are a set of $r+1$ $r$-dimensional vectors and $m$ and
$\tilde{\beta}$ are mass and coupling constant parameters.
It can be shown that the theories are integrable if
($\al_1,...,\al_r$) form the root system of a semi-simple classical
(rank~$r$) Lie algebra
$g_r$ (the longest root is taken to have length $\sqrt{2}$).  We
chose $\al_0 = -\sum_{j=1}^rn_j\al_j$
to be the negative of the
longest root (in case of $a_n^{(1)}$ all $n_j$ are equal to 1), such
that the system $(\al_0,...,\al_r)$ corresponds to the extended Dynkin
diagram of the affine Kac-Moody algebra $\hat{g}_r$. Thus, there is
one ATFT associated with each affine Kac-Moody algebra $\hat{g}_r$.

One has to distinguish between the two fundamentally different cases
of real and complex ATFT (i.e.\ with real or imaginary coupling
constant $\tilde{\beta}$).
In the case of real coupling constant, the Lagrangian defines a
unitary quantum field theory containing $r$ particles. Exact
S-matrices for the simply laced ATFTs, i.e.\ based on the $a,d$ and
$e$ series of Lie algebras, have been suggested by Braden et al.\ in
\cite{brade1} and independently by Christe and Mussardo in
\cite{chris}.
These
conjectures have also been shown to be consistent with the results of
perturbation theory to low orders in $\tilde{\beta}$.  S-matrices for
the non-simply laced ($b,c,f$ and $g$ algebras) as well as for the
twisted cases, in which the
poles corresponding to bound states need to have an additional
coupling constant dependence, have been
found by Delius et al.\ in \cite{deliu}.
Whereas the real case is fairly well understood, a complete
understanding of the complex ATFTs remains elusive.

The complex ATFTs (the coupling constant is purely imaginary:
$\tilde{\beta} = i \beta$ and $\beta$ real) are far more
complicated, but have also proved more interesting. Apart from complex
$a_1^{(1)}$~ATFT, which is the well known Sine-Gordon theory (SG),
these theories are in general non-unitary.
The most striking feature of complex ATFTs is the fact that they admit
classical soliton solutions. For the simplest case, the Sine-Gordon
theory, these soliton
solutions have been known for a long time. Only recently, however,
have soliton solutions been found for other ATFTs.
It can easily be seen that in the complex case the field is periodic
with respect to the weight lattice $\Lambda^*$ of $g_r$.
Since the original Lagrangian was chosen in a way that $\phi = 0$ was
the zero-energy solution, the constant fields $\phi =
\frac{2\pi}{\beta}\omega$ (with $\omega \in \Lambda^*$) all have zero
energy and thus, unlike in the
real case, the vacuum of the complex theory is degenerate.
This fact suggests already the existence of soliton solutions of the
classical equations of motion, which interpolate between different
vacua.
For each soliton solution the difference of the field between the two
vacua is called the topological charge $t$ of the solution. The
topological charge is also an element of the weight lattice
$\Lambda^*$ and can be written as:
\[
t = \frac{\beta}{2\pi}\int_{-\infty}^{\infty} dx \partial_x \phi
\hspace{7pt}, \hspace{7pt} \in \Lambda^*.
\]
Explicit expressions for one-soliton and multi-soliton solutions in
$a_n^{(1)}$ ATFT were
first found in \cite{hollo1} by using Hirota's method for
the solution of nonlinear differential equations.
Hollowood was also able to show that despite the complex form of the
Hamiltonian the solitons have real and positive energies and masses.
(For a more detailed discussion of the reality problem of affine Toda
solitons see references \cite{evans1},\cite{zhu2}.)
More one-soliton and multi-soliton solutions for other algebras have
been found in \cite{macka1},\cite{zhu1} and \cite{consta}. By using a
different method Olive et al.\ found a
general form for the soliton solutions for all ATFTs in \cite{olive}.

Apart from the elementary solitons there are also bound states of
solitons which can be seen as two solitons oscillating around a fixed
point. Whereas in SG theory only bound states with zero topological
charge (the `breathers') occur, in most other ATFTs there are
also bound states with non-zero topological charges (the `breathing
solitons' or `excited solitons'). In \cite{harde} classical
bound state solutions for $a_n^{(1)}$ were obtained by changing the
real velocity into an imaginary velocity in the expressions for the
two-soliton solutions.
Whereas in the real ATFTs only those $r$ particles exist which are
manifest in the Lagrangian, the complex ATFTs contain a rich spectrum
of solitons, bound states and excited solitons.  A great deal of work
has been done on the classical soliton solutions of complex affine
Toda field theories and their topological charges but heretofore very
little progress has been made towards a consistent quantization of
these solutions. In particular, the complete characterisation of the
bound-state spectrum for any other quantum ATFT, apart from
Sine-Gordon theory, has been an unsolved problem so far.

The layout of this paper is as follows: Section 2 gives a brief
introduction to the axioms
of exact S-matrix theory. The
Zamolodchikovs' solution of the Sine-Gordon model \cite{zamol1} and
Hollowood's S-matrix conjecture for the scattering of solitons in
$a_n^{(1)}$ ATFT \cite{hollo2} are reviewed. In section 3 we construct
bound states of two elementary solitons. Section 4 forms the main
part of this paper, the complete description of $a_2^{(1)}$
ATFT. In this section all
S-matrix elements involving the scattering of bound states are
constructed explicitly. It is also shown that the lowest
scalar bound states can be identified with the fundamental quantum
particles.
In section 5 the pole structure of these S-matrix elements
is discussed and a complete set of three point couplings is
conjectured.
Using these three-point couplings other poles are explained in terms
of higher order diagrams, some of which involve a generalized
Coleman-Thun mechanism.
Section 6 summarizes the results and
concludes with suggestions regarding possible further
directions of research.
\section{Exact Soliton S-Matrices}

In this section we briefly review the axioms of exact S-matrix theory
and describe some of the results from references \cite{zamol1}
and \cite{hollo2}.
\subsection{The Axioms of Exact S-Matrix Theory}

As mentioned in the introduction, the axioms of exact S-matrix theory
strongly constrain the possible form of the S-matrix of an integrable
theory. An integrable theory is one in which there exists an infinite
number of conserved charges. The existence of these charges implies
that any $N$-particle S-matrix `factorizes' into a
product of $\frac12 N(N-1)$ two-particle S-matrices. Moreover it can
be shown
(\cite{zamol1},
\cite{rajar}) that the S-matrices are purely elastic; that is, neither
particle creation nor destruction can occur.

Let us assume a theory with particles living in multiplets
$V_1,V_2$,...,$V_{n-1}$, which are complex vector spaces.
The two-particle S-matrix $S^{a,b}$ describing the
scattering process of two particles lying in $V_a$ and $V_b$
must act as an intertwiner on these spaces:
\[
S^{a,b}(\th_{ab}): V_a \otimes V_b \to V_b \otimes V_a.
\]
(Instead of using momenta, in two dimensions it is convenient
to work with the \mbox{rapidities $\th$} of particles, which
are given in terms of the two-momenta: $(p_0,p_1) =
m(\cosh|\th|,\sinh|\th|)$. The S-matrix is a function of the
rapidity difference $\th_{ab} = \th_a-\th_b$ of the incoming
particles.) \\
The S-matrix elements $S^{a,b}(\th)$ have to satisfy the following
conditions: \vspace{0.1cm} \\
(i) {\em Unitarity} \\
If $I_a$ denotes the identity operator on $V_a$, then
the unitary condition can be expressed as
\[
S^{a,b}(\th)S^{b,a}(-\th) = I_b \otimes I_a,
\]
(which does not imply that the Hamiltonian itself is
unitary). \vspace{0.1cm} \\
(ii) {\em Crossing Symmetry} \\
The matrix elements $S^{a,b}(\th)$ are symmetric under the
transformation
 $\th \to i\pi -\th$ such that
\be
S^{\ol a,b}(\th) = (I_b \otimes C_a)[\si S^{b,a}(i\pi-\th)]^{t_2}
\si (C_{\ol a} \otimes I_b), \label{crosssym}
\ee
in which $\ol a$ represents the conjugate of particle $a$.
For an exact definition of the charge conjugation operator $C_a$ and
the symbols $\si$ and $t_2$ see reference \cite{hollo2}.
\vspace{0.1cm} \\
(iii) {\em Analyticity} \\
$S^{a,b}(\th)$ is a meromorphic function of $\th$ with the only
singularities on the
physical strip $(0\leq \rm{Im}\th \leq \pi)$ at $\rm{Re}\th = 0$.
The `physical strip' in the $\th$-plane corresponds to the physical
sheet in the $s$-plane, where $s=(p_1+p_2)^2$ is just one of the usual
Mandelstam variables.
Simple poles in the physical strip correspond to bound states in the
direct or crossed channel. (In some cases simple poles could also
result from a process called the generalized Coleman-Thun mechanism,
which will be discussed in detail in section 5.) \vspace{0.1cm} \\
(iv) {\em Bootstrap} \\
If $S^{a,b}$ exhibits a simple pole $\th_{ab}^c$ on the physical strip
corresponding to a bound state in $V_c$ in the direct channel, then the
mass of this particle $c$ is given by the formula:
\be
m_c^2 = m_a^2 + m_b^2 + 2m_am_b\cos(\mbox{Im}\th_{ab}^c) \label{mass}
\ee
In this case there must also be
poles  $\th_{\ol ca}^{\ol b}$ in $S_{\ol ca}$ and $\th_{b\ol c}^{\ol
a}$  in $S_{b\ol c}$,
such that $\th_{ab}^c + \th_{\ol ca}^{\ol b} + \th_{b\ol c}^{\ol a} =
2\pi i$. The bootstrap equations express the fact that there is no
difference whether the scattering process with any particle in, say,
$V_d$ occurs before or after the fusion of particles  $a$ and $b$ into
particle $c$:
\be
[I_b \otimes S^{d,a}(\th-(i\pi-\th_{\ol ca}^{\ol b}))]
[S^{d,b}(\th+(i\pi-\th_{b\ol c}^{\ol a})) \otimes I_a] =
S^{d,c}(\th) \label{bootstrap}
\ee
\hspace{3cm}
(which is restricted to $V_d \otimes V_c \subset V_d \otimes V_a
\otimes V_b$). \vspace{0.1cm} \\
The bootstrap equations impose very strong constraints on the exact
form of the S-matrix elements and can be used to obtain higher
S-matrix elements from the lowest ones (see for instance
\cite{brade1}).
\subsection{Sine-Gordon Theory}

Only the simplest case of complex ATFTs, that of Sine-Gordon theory
($a_1^{(1)}$-ATFT),
has been solved completely by the Zamolodchikovs in \cite{zamol1},
where they give explicit expressions for all S-matrix elements
describing the scattering of solitons and their bound states.  They
used a noncommutative algebra to describe the full scattering
theory, which we will review briefly here.

Sine-Gordon theory contains two elementary solitons, the fundamental
soliton and its anti-soliton, which can be denoted as $A(\th)$ and
$\ol A(\th)$.
The possible scattering processes can then formally be written as:
\bea
A(\th_1)A(\th_2) &=& \tilde{S}^I(\th_{12}) A(\th_2)A(\th_1)
\nn \\
A(\th_1)\ol A(\th_2) &=& \tilde{S}^T(\th_{12}) \ol A(\th_2)A(\th_1) +
\tilde{S}^R(\th_{12}) A(\th_2)\ol A(\th_1) \label{zamolalg}
\eea
where $\tilde{S}^T(\th)$, $\tilde{S}^R(\th)$ and $\tilde{S}^I(\th)$
denote the S-matrix
elements for the transition, reflection and identical particle
processes respectively ($\th_{12} = \th_1-\th_2$ is the
rapidity difference of the incoming particles).
It should be noted that the reflection process is a purely quantum
mechanical effect, and that there is no reflection in the classical
scattering of soliton solutions (to discuss the reflection process on
the classical level one would need to consider complex time
trajectories, see e.g. \cite{hollo2} and references therein).

The above axioms of exact S-matrix theory can be used to obtain exact
expressions for the scattering amplitudes (see \cite{zamol1} and
references therein). However,
for a complete description of the SG theory one has to
augment
the algebra (\ref{zamolalg}) with symbols $B_p(\th)$ for the breather
states. If a breather $B_p$ corresponds to a pole $\Theta_p$ in
$S^T(\th)$ then the Zamolodchikovs formally defined:
\be
B_p(\frac{\th_1+\th_2}2) = \lim_{\th_2-\th_1\to \Theta_p}[A(\th_1)\ol
A(\th_2) \pm \ol A(\th_1)A(\th_2)] \label{sgbreather}
\ee
\hspace {5cm} ($+$ if $p$ is even, and $-$ if $p$ is odd).
\vspace{.2cm} \\
By considering triple products like $A(\th_1)\ol A(\th_2)A(\th_3)$ using
(\ref{zamolalg}), (\ref{sgbreather}) and taking the limit, one can
calculate explicit expressions of S-matrix elements for the scattering
of a breather with a fundamental soliton, or a breather with another
breather. (The complete Zamolodchikov algebra and expressions
for the S-matrix elements found by the Zamolodchikovs are given in
Appendix A.) In section 4 we will use this technique to calculate the
bound state scattering amplitudes for $a_2^{(1)}$ ATFT.
\subsection{A Quantum Group S-matrix for $a_2^{(1)}$-Solitons}

Quantum groups had not yet been introduced in the late seventies when
the complete SG S-matrix was found.
Only some years later was
it discovered that the S-matrix found in \cite{zamol1} has a quantum
group symmetry. The requirement of factorizability implies that the
two-particle
S-matrices satisfy a cubic identity called the `Yang-Baxter equation',
which plays an important role in the theory of quantum groups. Using
this fact, Bernard and LeClair in \cite{berna} expressed the SG
S-matrix in terms of the R-matrix of the quantum group
$U_q(Sl(2))$, which is the q-deformed universal enveloping algebra of
$Sl(2)$.
They also constructed non-local conserved charges for a variety of
massive integrable models. In particular they were able to show that
for an ATFT based on the algebra $g$ its non-local charges generate the
quantum loop algebra of $g^{\vee}$. This implies that the S-matrix for the
scattering of solitons, which is required to commute with the action
of these non-local charges, itself must display a quantum group
symmetry.
This quantum group symmetry of the S-matrix led Hollowood
to conjecture $U_q(Sl(n+1))$ symmetric S-matrices for the scattering
of solitons in \mbox{$a_n^{(1)}$ ATFTs} \cite{hollo2}. For our
purpose of
constructing the bound states of $a_2^{(1)}$ ATFT below, we will
restrict ourself to the description of Hollowood's construction
only for the case of $U_q(Sl(3))$.
(For a review of the theory of quantum groups and their
representations see for instance \cite{jimbo}, \cite{hollo2},
\cite{takht}.)

We will describe the construction of a soliton S-matrix in terms of
the $U_q(Sl(3))$ \mbox{R-matrix}, which is a trigonometric solution of
the Yang-Baxter-Equation (YBE). In its general form this equation can
be written as:
\bea
[\cR^{U_2,U_3}(x)\otimes I_{U_1}] [I_{U_2}\otimes
\cR^{U_1,U_3}(xy)] [\cR^{U_1,U_2}(y)\otimes I_{U_3}] =
\nn \\
= [I_{U_3}\otimes\cR^{U_1,U_2}(y)]
[\cR^{U_1,U_3}(xy)\otimes  I_{U_2}] [I_{U_1}\otimes
\cR^{U_2,U_3}(x)] \label{YBE}
\eea
where $U_1$, $U_2$ and $U_3$ are complex vector spaces,
$\cR^{U_i,U_j}: U_i\otimes U_j \to U_j\otimes U_i$ are linear
maps and $I_{U_i}$ is the identity on $U_i$, thus both sides of
(\ref{YBE}) map
$U_1\otimes U_2 \otimes U_3$ into $U_3\otimes U_2 \otimes U_1$
(x is called the spectral parameter).

We are interested in representations associated with the
two fundamental representations $(\rho_a,V_a) \hsv (a=1,2)$ of $a_2$.
(The commonly used notations for $\rho_1$ and $\rho_2$ are \mbox{$\ul
3$ and $\ol{\ul 3}$}).
Here we can take $V_1 \cong C^3$ and choose $(e_1,e_2,e_3)$ as a basis of
$V_1$. As shown in \cite{jimbo} a basic R-matrix can then be
expressed in terms of the generator $T$ of the Hecke-algebra ${\cal
H}_2$ (which is a generalisation of the symmetric group ${\cal S}_2$):
\[
\cR^{1,1}(x) = xT^{-1} - x^{-1}T
\]
and we can use the particular representation
\vspace{3pt}

\hspace{3cm} $T(e_i\otimes e_j) = \left\{
\begin{array}{ll} q^{-1}e_i\otimes e_j & \mbox{if $i=j$} \\
(q^{-1}-q)e_i\otimes e_j + e_j\otimes e_i & \mbox{if $i>j$} \\
e_j\otimes e_i & \mbox{if $i<j$}.
\end{array}
\right. $ \vspace{0.3cm} \\
It can easily be confirmed that $\cR^{1,1}(x) \in
\mbox{End}(V_1\otimes V_1)$ satisfies the YBE
for the special case $U_1=U_2=U_3=V_1$.
If we define the basic matrices $E_{ij}$ acting on $V_1$ by $E_{ij}e_k
\equiv \delta_{jk}e_i$ (for $i,j,k = 1,2,3$) then we can write
explicitly:
\bea
\cR^{1,1}(x) &=& (xq-x^{-1}q^{-1}) \sum_{i=1}^3 E_{ii}\otimes
E_{ii} + x(q-q^{-1}) \sum_{i<j} E_{ii}\otimes E_{jj} \nn \\
&& + x^{-1}(q-q^{-1}) \sum_{i>j} E_{ii}\otimes E_{jj} + (x-x^{-1})
\sum_{i\neq j} E_{ji}\otimes E_{ij}. \label{rmatrix}
\eea
$\cR^{1,1}$ can also be expressed conveniently in terms of the
projection operators \mbox{$s^{\pm}_2 = \frac 1{[2]_q}(q^{\pm 1}\pm
T)$}, which are the quantum group analogues of the full symmetrizer
(antisymmetrizer) and $[n]_q = (q^n-q^{-n})/(q-q^{-1})$,
\be
\cR^{1,1}(x) = (xq-x^{-1}q^{-1})s_2^+ + (x^{-1}q-xq^{-1})s_2^-.
\label{specdecom}
\ee
This formula is called the spectral decomposition of
$\cR^{1,1}$.
By applying the fusion procedure \cite{jimbo}, we
can construct
other solutions of the YBE from the basic solution $\cR^{1,1}(x)$ in
the following way. The module $V_2$ of the second fundamental
representation can be defined as subspace of the tensor product of
$V_1$ with itself by using projector $s_2^-$:
\[
V_2 \equiv s_2^-(V_1 \otimes V_1)
\]
and we define
\be
\cR^{1,2}(x) = [I\otimes \cR^{1,1}(x(-q)^{\frac12})]
[\cR^{1,1}(x(-q)^{-\frac12}) \otimes  I] \label{fusion1}
\ee
which acts on $V_1 \otimes V_2$ ($I$ denotes the identity on $V_1$).
In the same way we can define more R-matrices:
\bea
\cR^{2,1}(x) &=& [\cR^{1,1}(x(-q)^{\frac12})\otimes I]
[I \otimes \cR^{1,1}(x(-q)^{-\frac12})] \nn \\
\cR^{2,2}(x) &=& [I\otimes \cR^{2,1}(x(-q)^{\frac12})]
[\cR^{2,1}(x(-q)^{-\frac12}) \otimes  I]. \label{fusion2}
\eea
Hollowood proved in \cite{hollo2} (in which this method was applied
for the more general case of $Sl(n)$) that these definitions indeed
yield new solutions of the YBE. The fusion procedure for the
construction of higher R-matrices can be understood as the
analogue for obtaining new irreducible representations of
classical Lie algebras by considering spectral decompositions of
tensor products of the fundamental representations.

$a_2^{(1)}$ ATFT contains three (mass degenerate) solitons, which
all lie in the same multiplet $V_1$ transforming under the fundamental
representation $\rho_1$. Analogously, their corresponding
antisolitons transform under $(\rho_2,V_2)$.
Hollowood defined the following S-matrix, describing the scattering of
these fundamental solitons and antisolitons,
\be
S^{a,b}(\th) \equiv X^{a,b}(x(\th))\cR^{a,b}(x(\th)) \hspace{2cm}
(a,b=1,2), \label{smatrix}
\ee
in which the connections between the deformation parameter $q$, the
coupling constant $\beta$, the rapidity difference $\th$ and the
spectral parameter $x$ are given by:
\bea
q &=& - e^{-i\pi\la} \hsv ,\mbox{where} \hsv \la =
\frac{4\pi}{\beta^2}-1  \nn  \\
x &=& e^{-i\pi\mu} \hspace{16pt} ,\mbox{where} \hsv \mu =
i\frac{3\la}{2\pi}\th. \label{connect}
\eea
(The form of the function $\la$ has first been suggested by
Arinshtein, Fateev and Zamolodchikov in \cite{arins}.)
The scalar factor $X^{a,b}$ in (\ref{smatrix}) ensures crossing
symmetry and was
given explicitly in terms of Gamma-functions in \cite{hollo2}.
Hollowood was also able to give further evidence for this S-matrix by
showing that the results from semiclassical calculations are in
agreement with his S-matrix conjecture.
Starting from this soliton S-matrix, in \mbox{section 4} we will
construct all S-matrix elements involving the scattering of bound
states.

One readily discovers a similarity in the form of equations
(\ref{fusion1}) and (\ref{fusion2}) with the form of the bootstrap
equations (\ref{bootstrap}). And indeed the fusion procedure to
construct new solutions of the YBE is essentially the same procedure
as constructing higher S-matrices from lower ones by applying the
bootstrap equations. In the following section we will again make use
of the fusion procedure by constructing bound
states, which transform under irreducible representations in the
tensor product of the two fundamental representations.
In order to do this we will use a non-commutative algebra, as the
Zamolodchikovs did for SG theory.
\section{The Bound States}

In this section we will construct explicit expressions of all possible
bound states of two elementary solitons or antisolitons in $a_2^{(1)}$
ATFT (see also \cite{hollo2}).

As already mentioned above,
$a_2^{(1)}$ ATFT contains three fundamental solitons and their
corresponding antisolitons, which we will denote as $A^j$ and $\ol
A^j$ ($j=1,2,3$) respectively. They are mass degenerate and their
classical masses are $m_A = 2m\sin(\frac{\pi}3)$.
We will formally define a Zamolodchikov
algebra in order to apply the same method the Zamolodchikovs used for
SG theory.  Due to the unconventional action of the charge
conjugation operator defined in \cite{hollo2} it is necessary to
redefine the in and out states in order to obtain explicit crossing
symmetric expressions for the amplitudes of the transition, reflection
and identical particle processes.
In terms of the basis elements $e_i$ ($i=1,2,3$) of $V_1$ we redefine
the in and out states as follows:
\be
A^j(\th_1)A^k(\th_2)= e^{(j-k)\frac12\la\th_{12}}e_j\otimes e_k
\hspace{2cm} (j,k = 1,2,3). \label{redef}
\ee
$A^j(\th_1)A^k(\th_2)$ denotes two incoming solitons (of
species j and k) if $\th_1 > \th_2$ and two outgoing solitons if
$\th_2 > \th_1$.
(This redefinition of
states corresponds to a change from the homogeneous gradation to the
principal gradation of the underlying charge algebra, as described in
\cite{brack}, and the same
result could therefore equally be obtained by a suitable gauge
transformation, as done in \cite{berna} for SG.)
With this definition of the symbols $A^j(\th)$ we can formally write
the scattering processes of fundamental solitons in form of the
following commutation relations:
\bea
A^j(\th_1)A^j(\th_2) &=& S^I(\th_{12}) A^j(\th_2)A^j(\th_1) \nn
\\
A^j(\th_1)A^k(\th_2) &=& S^T(\th_{12}) A^k(\th_2)A^j(\th_1) +
S^{R(j,k)}(\th_{12}) A^j(\th_2)A^k(\th_1) \label{commrel}
\eea
in which $S^T$, $S^{R(j,k)}$ and $S^I$ are scalar functions of one
complex variable, which describe the transition, reflection and
identical particle processes of fundamental solitons respectively.
(The notation
$S^{R(j,k)}$ indicates that the reflection amplitude, unlike in the SG
case, depends on the species $j,k$ of the two solitons.)
In order to obtain these amplitudes from the quantum group S-matrix
(\ref{smatrix}) we use the explicit expression (\ref{rmatrix}) and let
$S^{1,1}$ formally act on the in-states (\ref{redef}):
\bea
S^{1,1}(\th_{12})[A^j(\th_1)A^j(\th_2)] &\equiv&
S^I(\th_{12})A^j(\th_2)A^j(\th_1) \nn \\
S^{1,1}(\th_{12})[A^j(\th_1)A^k(\th_2)] &\equiv&
S^T(\th_{12})A^k(\th_2)A^j(\th_1) +
S^{R(j,k)}(\th_{12})A^j(\th_2)A^k(\th_1). \nn
\eea
 From this definition, using the relations (\ref{connect}), we
obtain\footnote {These amplitudes coincide with those obtained by
Nakatsu in \cite{nakat} modulo some misprints.}:
\bea
S^I(\th) & = & 2i \sin(i\frac 32\la\th + \la\pi)f_{0,0}(\mu)
\nn \\
S^T(\th) & = & 2i
\sin(-i\frac32\la\th) f_{0,0}(\mu) \nn \\
S^{R(j,k)}(\th) & = & 2i
\sin(\la\pi)e^{\nu(j,k)\frac12\la\th}f_{0,0}(\mu)
\eea
\vspace{3pt}
\hspace{2.5cm} in which $ \nu(j,k) \equiv \left\{
\begin{array}{ll}+1
& \mbox{if (j,k) = (1,2),(2,3) or (3,1)} \\ -1 &
\mbox{if (j,k) = (2,1),(3,2) or (1,3)} .
\end{array}
\right.$ \\
Our motivation for the change of notation for the scalar factor
(our $f_{0,0}(\mu)$ is identical to $X^{1,1}(x(\th))$ in Hollowood's
notation) will become clearer below.
The scalar factor $f_{0,0}(\mu)$ is given explicitly in section 4.

In order to obtain explicit expressions for the scattering amplitudes
for processes involving antisolitons, in
the case of $a_2^{(1)}$ it is sufficient to know the
explicit expression (\ref{rmatrix}) for $\cR^{1,1}$.
Instead of using the fusion procedure (or equivalently the bootstrap
equations) to obtain the higher S-matrices $S^{1,2}(\th)$ and
$S^{2,2}(\th)$, we can get all the information about the scattering
amplitudes by applying the crossing symmetry relation (equation
(\ref{crosssym})). Simply taking the crossed versions of the processes
in (\ref{commrel}), we obtain the following commutation relations:
\bea
A^j(\th_1)\ol A^k(\th_2) &=& S^T(i\pi-\th_{12})\ol
A^k(\th_2)A^j(\th_1) \nn \\
A^j(\th_1)\ol A^j(\th_2) &=& S^I(i\pi-\th_{12})\ol
{A}^j(\th_2)A^j(\th_1) +
S^{R(k,j)}(i\pi-\th_{12}) \ol A^k(\th_2)A^k(\th_1) + \nn \\
&&+ S^{R(l,j)}(i\pi-\th_{12}) \ol A^l(\th_2)A^l(\th_1)
\label{ancommrel}
\eea
The relations for $\ol A-\ol A$ scattering processes are
identical to relations (\ref{commrel}).

Now we are able to formally construct bound states of fundamental
solitons, using the same procedure as the Zamolodchikovs in
\cite{zamol1}.
The axioms of exact S-matrix theory state that bound states
correspond to simple poles on the physical strip (i.e. $0 \leq
\mbox{Im}(\th) \leq \pi$, Re($\th) = 0$) in the direct channel of the
S-matrix. \vspace{.2cm} \\
\ul{$S^I(\th)$} contains the following poles:
\be
\th = i\frac{2\pi}{3\la}p \hspace{2cm} (\mbox{for } p=1,2,...\leq
\frac32\la), \label{brpole}
\ee
which correspond to scalar bound states (bound states with zero
topological charge, the so-called breathers) in the cross channel.
Since the topological charge has to be conserved, these breathers must
be bound states of a soliton-antisoliton pair of the same species.
We have to add symbols $B_p$ for breathers and $\ol B_p$ for their
conjugate breathers to the algebra (\ref{commrel}). Since these bound
states are required to transform under irreducible representations,
which appear in the spectral decomposition of the tensor product of
$\rho_1$ with $\rho_2$, they must be defined as:
\bea
B_p(\frac{\th_1+\th_2}2)= \lim_{\th_2-\th_1\to
-i\frac{2\pi}{3\la}p+i\pi}[\sum_{m=1}^3\al_m^{(p)}A^m(\th_1)\ol
A^m(\th_2)] \nn \\
\ol B_p(\frac{\th_1+\th_2}2)= \lim_{\th_2-\th_1\to
-i\frac{2\pi}{3\la}p+i\pi}[\sum_{m=1}^3\al_m^{(p)}\ol A^m(\th_1)
A^m(\th_2)] \label{brdef}
\eea
\hspace{4cm} in which $p=1,2,...\leq \frac32\la$, and
\[
(\al_1^{(p)},\al_2^{(p)},\al_3^{(p)}) = \left\{
\begin{array}{ll} \eta(e^{i\frac{2\pi}3},1,e^{-i\frac{2\pi}3}) &
\mbox{if p = 1,4,7,...} \\
\eta(e^{-i\frac{2\pi}3},1,e^{i\frac{2\pi}3}) &
\mbox{if p = 2,5,8,...} \\ \eta(1,1,1) & \mbox{if p = 3,6,9,...}
\end{array}
\right. \nn
\]
($\eta$ is an arbitrary phase factor.)
\vspace{0.2cm} \\
The transition amplitude \ul{$S^T(\th)$} contains the following poles:
\be
\th = -i\frac{2\pi}{3\la}p + i\frac{2\pi}3 \hspace{2cm} (\mbox{for }
p=0,1,...\leq \la).
\label{exsolpole}
\ee
These poles correspond to bound states with non-zero topological
charges, which we will call `excited solitons'. We can define symbols
$A_p^j(\th)$ and $\ol A_p^j(\th)$
for these excited solitons as:
\bea
A^j_p(\frac{\th_1+\th_2}2) = \lim_{\th_2-\th_1\to -i
\frac{2\pi}{3\la}p + i \frac{2\pi}3} \eta [\ga_1 \ol A^k(\th_1) \ol
A^l(\th_2)  + \ga_2
\ol A^l(\th_1)\ol A^k(\th_2)] \nn \\
\ol A^j_p(\frac{\th_1+\th_2}2) = \lim_{\th_2-\th_1\to -i
\frac{2\pi}{3\la}p + i \frac{2\pi}3} \eta [\ga_1
A^k(\th_1)A^l(\th_2) + \ga_2
A^l(\th_1)A^k(\th_2)] \label{exsoldef}
\eea
for $p=0,1,...\leq \la$ and $\{j,k,l\} = \{1,2,3\}$.
We have to take $\ga_1/\ga_2 = (-1)^pe^{i\frac{\pi}3(\la-p)}$
(with an arbitrary phase factor $\eta$).
The form of the parameters $\ga_j$ in the
definition of $A_p^j$ is determined by the requirement
that the bound states transform under the quantum analogues of the
irreducible representations, appearing in the tensor product of the
fundamental representations $\rho_1$ or $\rho_2$ with themselves.
Under which particular representation a given bound state transforms
can be determined by the position of its corresponding pole in the
spectral decomposition of the S-matrix.
Considering the spectral decomposition
(\ref{specdecom}), we see that the S-matrix poles, corresponding to
excited solitons, appear in the factor in front of the projector
$s_2^-$.
This implies that the excited solitons $A_p^j$ (antisolitons
$\ol A_p^j$) must transform under the quantum group analogue of the
fundamental representation $\rho_1$ ($\rho_2$). This requirement fixes
the $\ga_j$ uniquely up to a phase factor $\eta$.
To clarify this let us consider the example of $\ol
A_p^1$: Using (\ref{exsoldef}) and the redefinition (\ref{redef}) we
see that with the given relation for $\ga_1$ and $\ga_2$ we have:
\[
\ol A_p^1(\th) \sim \frac1{q+q^{-1}}(e_2\otimes e_3 - qe_3\otimes e_2).
\]
It can easily be proved that the expression on the right-hand side is
indeed invariant under the action of the projector $s_2^-$, which was
defined in section 2.3.
The fact that $A_p^j(\th)$ and $\ol A_p^j(\th)$ transform under the
same representations as the fundamental solitons $A^j(\th)$ and $\ol
A^j(\th)$ further justifies the term `excited
solitons'. Note that for $p=0$ these states are in fact identical to
the elementary solitons and antisolitons.

Yet another important point to mention here is the fact that the poles
corresponding to the excited solitons only appear in $S^T(\th)$ and
not in $S^I(\th)$. This confirms the classical result obtained in
\cite{harde} that there are no bound states of two solitons of the
same species (i.e. \ul{no} $A^j-A^j$ bound states) in the theory.
There are no other poles apart from (\ref{brpole}) and
(\ref{exsolpole}) in the soliton S-matrix, and so (\ref{brdef}) and
(\ref{exsoldef}) give all possible bound states of two elementary
solitons in $a_2^{(1)}$ ATFT.
\section{Scattering of Bound States}

The formal definitions of symbols $A_p^j$ and $B_p$ correspond
to the calculation of the appropriate residues of the S-matrix
elements and therefore enable us to calculate all bound state S-matrix
elements solely by using the elementary algebra
(\ref{commrel}),(\ref{ancommrel}) in the
same way as done by the Zamolodchikovs for Sine-Gordon.
This can be a rather elaborate procedure, for which we will only show
one example in detail.

In order to obtain the scattering amplitude for the scattering of an
elementary soliton $A^j$ with a breather $B_p$ we need to consider the
triple product $A^j(\th_1)\sum_{m=1}^3[\al_m^{(p)}A^m(\th_2)\ol
A^m(\th_3)]$. Using the exchange relations
(\ref{commrel}) and (\ref{ancommrel}) and taking the limit
\mbox{($\th_3-\th_2\to -i\frac{2\pi}{3\la}p + i\pi)$} we obtain that
the scattering process is
purely a transition process, such that we have to add to the
Zamolodchikov algebra the exchange relation:
\[
A^j(\th_1)B_p(\th_2) = S_{AB_p}(\th_{12})B_p(\th_2)A^j(\th_1).
\]
The details of this calculation appear in Appendix B. There the
transition amplitude for this process has been calculated to be
\[
S_{AB_p}(\th) = S^T(\th-i\frac{\pi p}{3\la} + i\frac{\pi}2)
S^T(-\th-i\frac{\pi p}{3\la} + i\frac{3\pi}2)
\]
By using the properties of the scalar factor $f(\mu)$ derived in
Appendix C, we can simplify the expression for $S_{AB_p}$ and end up
with only a finite product of sine-functions:
\be
S_{AB_p}(\th) = \prod_{n=1}^p \frac{\sin(\frac{\th}{2i} +
\frac{\pi}{6\la}(p-2n)+\frac{\pi}4)
\sin(\frac{\th}{2i} +
\frac{\pi}{6\la}(p-2(n-1))+\frac{\pi}4)}
{\sin(\frac{\th}{2i} +
\frac{\pi}{6\la}(p-2(n-1))+\frac7{12}\pi)
\sin(\frac{\th}{2i} +
\frac{\pi}{6\la}(p-2n)-\frac{\pi}{12})}.
\ee
By starting with the triple product
$\sum_{m=1}^3[\al_m^{(p)}A^m(\th_1)\ol A^m(\th_2)]A^j(\th_3)$ and
taking the limit ($\th_2-\th_1\to -i\frac{2\pi}{3\la}p + i\pi)$, in
the same way
we obtain the amplitude for the transition process:
\[
B_p(\th_1)A^j(\th_2) = S_{B_pA}(\th_{12})A^j(\th_2)B_p(\th_1)
\]
in which
\be
S_{B_pA}(\th) = \prod_{n=1}^p \frac{\sin(\frac{\th}{2i} +
\frac{\pi}{6\la}(p-2n)-\frac7{12}\pi)
\sin(\frac{\th}{2i} +
\frac{\pi}{6\la}(p-2(n-1))+\frac{\pi}{12})}
{\sin(\frac{\th}{2i} +
\frac{\pi}{6\la}(p-2(n-1))-\frac{\pi}4)
\sin(\frac{\th}{2i} +
\frac{\pi}{6\la}(p-2n)-\frac{\pi}4)}.
\ee
\vspace{.2cm}

We would like to draw attention to the fact that $S_{AB_p}(\th) \neq
S_{B_pA}(\th)$ which initially seems to be in
violation of left-right symmetry. However, closer examination reveals
that this phenomenon is closely related to the fact that the breathers
of $a_2^{(1)}$-ATFT are not self-conjugate (i.e. $B_p \neq \ol B_p$),
as in SG-theory. The breathers $B_p$ and $\ol B_p$ are indeed two
different
kinds of particles despite the fact that both have the same mass and
transform under the singlet representation. It is presently not clear
to us whether they have different higher spin conserved local charges,
like the mass degenerate particles in real ATFT (see \cite{brade1}), or
whether there are any non-local charges which serve to distinguish
them uniquely.

Writing the breather states as bound states of
fundamental solitons, {\em Figure 1} shows an $B_p - A^j$ scattering
process on the left hand side and the same process with reversed
ordering and reversed signs of the rapidities of the incoming
particles on the right hand side. This
illustrates the fact that the requirement of left-right
symmetry implies $S_{AB_p}(\th) = S_{\ol B_pA}(\th)$ and
\underline{not} $S_{AB_p}(\th) = S_{B_pA}(\th)$.
It can be seen easily that $S_{AB_p}$ and $S_{B_pA}$ are
each separately crossing symmetric (i.e. $S_{AB_p}(i\pi-\th) =
S_{AB_p}(\th)$) and therefore satisfy the required symmetry
conditions:
\bea
S_{AB_p}(\th) = S_{\ol B_pA}(\th) = S_{\ol A \hspace{1pt} \ol
B_p}(\th) = S_{B_p\ol A}(\th); \nn \\
S_{B_pA}(\th) = S_{A\ol B_p}(\th) = S_{\ol B_p\ol A}(\th) = S_{\ol
AB_p}(\th).
\eea

\vspace{1.5cm}
%
%
%%%% Left-Right Symmetry in A-Bp Scattering  %%%%
%
\begin{center}
\begin{picture}(300,140)(0,-50)
\put(150,0){\line(0,1){10}}
\put(150,20){\line(0,1){10}}
\put(150,40){\line(0,1){10}}
\put(150,60){\line(0,1){10}}
\put(150,80){\line(0,1){10}}
\put(150,100){\line(0,1){10}}
\put(150,120){\line(0,1){10}}
\put(120,0){\line(-1,2){60}}
\put(30,0){\line(1,1){40}}
\put(60,0){\line(1,4){10}}
\put(70,40){\line(1,2){40}}
\put(85,70){\circle*{8}}
\put(180,0){\line(1,2){60}}
\put(270,0){\line(-1,1){40}}
\put(240,0){\line(-1,4){10}}
\put(230,40){\line(-1,2){40}}
\put(215,70){\circle*{8}}
\put(25,-10){\shs{\fns{$\ol A_j$}}}
\put(55,-10){\shs{\fns{$A_j$}}}
\put(115,-10){\shs{\fns{$A_k$}}}
\put(63,52){\shs{\fns{$B_p$}}}
\put(55,122){\shs{\fns{$A_k$}}}
\put(105,122){\shs{\fns{$B_p$}}}
\put(265,-10){\shs{\fns{$\ol A_j$}}}
\put(235,-10){\shs{\fns{$A_j$}}}
\put(175,-10){\shs{\fns{$A_k$}}}
\put(226,52){\shs{\fns{$\ol B_p$}}}
\put(235,122){\shs{\fns{$A_k$}}}
\put(185,122){\shs{\fns{$\ol B_p$}}}
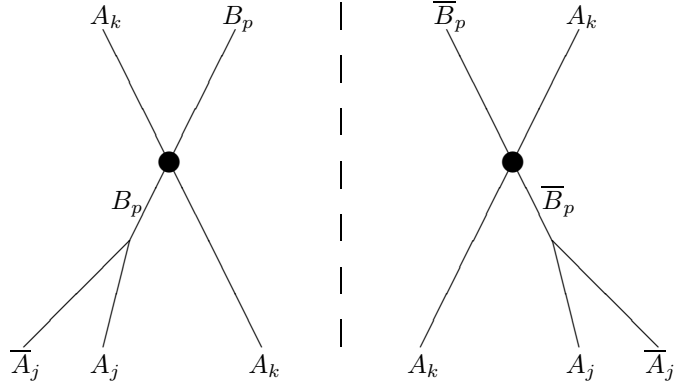
\put(40,-40){\shs{\em Figure 1: Left-right symmetry in $S_{B_pA}$}}
\end{picture}
\end{center}

\vspace{0.5cm}
By further applying the same method we constructed all S-matrix
elements for the scattering of all possible bound states. Without
going into further detail of the calculations we will list the
results here. We will use the following notations
and abbreviations, some of which have been
borrowed from references \cite{brade1} and \cite{hollo2}:
\[
\mu \equiv i\frac{3\la}{2\pi}\th, \hspace{5pt} \rm{and} \hspace{5pt}
\la \equiv \frac{4\pi}{\beta^2} - 1,
\]
\[
\lb x \rb \equiv sin(\frac{\th}{2i} + \frac{\pi x}{6\la})
\]
\\
\be
\bigl( x \bigr) \equiv \frac{sin(\frac{\th}{2i}+\frac{\pi x}{6\la})}
{sin(\frac{\th}{2i}-\frac{\pi x}{6\la})} = \frac{\lb x \rb}{\lb -x
\rb}
\ee

\vspace{0.4cm}
\hspace{2cm} and $ \nu(j,k) \equiv \left\{
\begin{array}{ll}+1
& \mbox{if (j,k) = (1,2),(2,3) or (3,1)} \\ -1 &
\mbox{if (j,k) = (2,1),(3,2) or (1,3)} .
\end{array}
\right. $
\newpage
$a_2^{(1)}$ affine Toda field theory contains the following kinds of
solitons, antisolitons and bound states:\\
\begin{tabbing}
a) {\em Fundamental solitons}: \hspace{20pt} \= $A^j(\th)$
\hspace{40pt} \= $(j=1,2,3)$ \\
b) {\em Fundamental antisolitons}: \> $\ol A^j(\th)$ \> $(j=1,2,3)$ \\
c) {\em Breathers and conjugate breathers ($A^j-\ol A^j$ bound
states)}: \\
\> $B_p(\th), \ol B_p(\th)$ \> $(p=1,2,3,... \leq \frac32 \la)$ \\
d) {\em Excited solitons and antisolitons ($A^l-A^k$ bound states)}:
\\
\> $A^j_p(\th), \ol A^j_p(\th)$ \> $(j=1,2,3; \hspace{4pt} p=0,1,2,...
\leq\la)$ \\
\end{tabbing}
Using these symbols all possible scattering processes can formally be
written in terms of the following commutation relations:
\vspace{0.6cm} \\
{\bf a) soliton-soliton scattering:}
\bea
A^j(\th_1)A^j(\th_2) &=& S^I(\th_{12}) A^j(\th_2)A^j(\th_1) \nn
\\
A^j(\th_1)A^k(\th_2) &=& S^T(\th_{12}) A^k(\th_2)A^j(\th_1) +
S^{R(j,k)}(\th_{12}) A^j(\th_2)A^k(\th_1) \nn
\eea
{\bf b) soliton-breather scattering:}
\bea
A^j(\th_1)B_p(\th_2) = S_{AB_p}(\th_{12}) B_p(\th_2)A^j(\th_1)
\nn \\
B_p(\th_1)A^j(\th_2) = S_{B_pA}(\th_{12}) A^j(\th_2)B_p(\th_1)
\nn
\eea
{\bf c) breather-breather scattering:}
\[
B_r(\th_1)B_p(\th_2) = S_{B_rB_p}(\th_{12}) B_p(\th_2)B_r(\th_1)
\]
{\bf d) excited soliton-breather scattering:}
\bea
A^j_r(\th_1)B_p(\th_2) = S_{A_rB_p}(\th_{12}) B_p(\th_2)A^j_r(\th_1)
\nn \\
B_p(\th_1)A^j_r(\th_2) = S_{B_pA_r}(\th_{12}) A^j_r(\th_2)B_p(\th_1)
\nn
\eea
{\bf e) excited soliton-excited soliton scattering:}
\bea
A^j_p(\th_1)A^j_r(\th_2) &=& S^I_{p,r}(\th_{12})
A^j_r(\th_2)A^j_p(\th_1) \nn \\
A^j_p(\th_1)A^k_r(\th_2) &=& S^T_{p,r}(\th_{12})
A^k_r(\th_2)A^j_p(\th_1) +
S^{R(j,k)}_{p,r}(\th_{12}) A^j_r(\th_2)A^k_p(\th_1) \nn
\eea
\\
Relations a) and b) can be regarded as special cases of d) and e).
\newpage
{\bf The S-matrix elements are given by:}\\
\bea
S^I(\th) & = & 2i \sin(i\frac 32\la\th + \la\pi)f_{0,0}(\mu)
\nn \\
S^T(\th) & = & 2i
\sin(-i\frac32\la\th) f_{0,0}(\mu) \nn \\
S^{R(j,k)}(\th) &
= & 2i \sin(\la\pi)e^{\nu(j,k)\frac12\la\th}f_{0,0}(\mu)
\label{solsol}
\eea
\bea
S_{AB_p}(\th) = \prod_{n=1}^p\frac{\lb p-2n+\frac32 \la\rb \lb p-2(n-1)
+\frac32 \la\rb} {\lb p-2(n-1)+\frac72 \la \rb \lb p-2n -\frac12 \la\rb}
\nn \\
S_{B_pA}(\th) = \prod_{n=1}^p\frac{\lb p-2n-\frac72 \la\rb \lb p-2(n-1)
+\frac12 \la\rb} {\lb p-2(n-1)-\frac32 \la\rb \lb p-2n -\frac32 \la\rb}
\label{solbr}
\eea
\bea
S_{B_rB_p}(\th) &=&
 \prod_{n=1}^p\biggl(r+p-2n\biggr)\biggl(r+p-2(n-1)\biggr) \nn
\\
&&\times\biggl(p-r-2n-2\la\biggr)\biggl(p-r-2(n-1)+2\la\biggr)
\label{brbr}
\eea
\bea
S_{A_rB_p}(\th) = \prod_{n=1}^p\frac{\lb r+p-2n+\frac32 \la\rb \lb
r+p-2(n-1) -\frac12 \la\rb}
{\lb r+p-2(n-1)+\frac72\la\rb \lb r+p-2n +\frac72 \la\rb} \nn \\
\times \frac{ \lb p-r-2n+\frac72\la\rb \lb p-r-2(n-1)+\frac32\la\rb}
{\lb p-r-2(n-1)-\frac12\la\rb \lb p-r-2n-\frac12\la\rb} \nn \\
S_{B_pA_r}(\th) = \prod_{n=1}^p\frac{\lb r+p-2n+\frac12 \la\rb \lb
r+p-2(n-1) +\frac12 \la\rb} {\lb r+p-2(n-1)+\frac52\la\rb \lb r+p-2n
-\frac32 \la\rb} \nn \\
\times \frac{\lb p-r-2n-\frac72\la\rb \lb p-r-2(n-1)+\frac52\la\rb}
{\lb p-r-2(n-1)-\frac32\la\rb \lb p-r-2n+\frac12\la\rb} \label{exsolbr}
\eea
\bea
S^I_{p,r}(\th) & = & 2i \sin(i\frac 32\la\th - \frac{\pi(p+r)}2 +
\la\pi)f_{p,r}(\mu) \nn \\
S^T_{p,r}(\th) & = & 2i
\sin(-i\frac32\la\th + \frac{\pi(p+r)}2) f_{p,r}(\mu) \nn \\
S^{R(j,k)}_{p,r}(\th) & = & 2i \sin(\la\pi)e^{\nu(j,k)(\frac12\la\th
- i \frac{\pi(r+p)}2)} f_{p,r}(\mu) \label{exsolexsol}
\eea
\vspace{0.5cm} \\
For the sake of simplicity we only listed the relations for
solitons and breathers, since all processes involving antisolitons or
conjugate breathers can easily be obtained from these by taking their
crossed version (i.e. $\th \to i\pi -\th$). For example the breather
- conjugate breather scattering process is described by:
\[
B_p(\th_1)\ol B_r(\th_2) = S_{B_p\ol B_r}(\th_{12}) \ol B_r(\th_2)
B_p(\th_1)
\]
in which
\bea
S_{B_p\ol B_r}(\th) = S_{B_rB_p}(i\pi-\th) &=&
\prod_{n=1}^p \biggl(p-r-2n+3\la\biggr)
\biggl(p-r-2(n-1)+3\la\biggr)\nn \\
&&\times\biggl(r+p-2(n-1)+5\la\biggr)\biggl(r+p-2n+\la\biggr)
\eea
It can easily be proven that these breather-breather scattering
amplitudes satisfy the obvious symmetry conditions:
\bea
S_{B_rB_p}(\th) = S_{B_pB_r}(\th) = S_{\ol B_r\ol B_p}(\th) = S_{\ol
B_p\ol B_r}(\th); \nn  \vspace{0.1cm} \\
S_{\ol B_pB_r}(\th) = S_{B_r\ol B_p}(\th) = S_{\ol B_rB_p}(\th) =
S_{B_p\ol B_r}(\th). \nn
\eea
\vspace{0.1cm}

All the scalar factors in the expressions (\ref{solsol}) and
(\ref{exsolexsol}) can be expressed in terms of the fundamental
function $f(\mu)$ given by Hollowood:
\bea
f(\mu) = \frac1{2i \sin(\pi(\mu+\lambda))} \prod_{j=1}^{\infty}
\frac{\Gamma[1+\mu+(3j-3)\lambda] \Gamma[\mu+3j\lambda]}
{\Gamma[1-\mu+(3j-3)\lambda] \Gamma[-\mu+3j\lambda]} \nn \\
\times\frac{\Gamma[-\mu+(3j-2)\lambda] \Gamma[1-\mu+(3j-1)\lambda]}
{\Gamma[\mu+(3j-2)\lambda] \Gamma[1+\mu+(3j-1)\lambda]}.
\eea
\\
In Appendix C we show the calculations which led to the following
expression of $f_{p,r}$:
\bea
f_{p,r}(\mu) =& \biggl(r+p\biggr)
\biggl(-r-p-2\la\biggr) \biggl(p-r+2\la\biggr)
\biggl(r-p+2\la\biggr) \nn \\
&\times\prod_{n=1}^r \frac{\lb r+p-2(n-1)+2\la\rb \lb r+p-2n\rb}{\lb
r+p-2n-2\la\rb \lb r+p-2(n-1)-2\la\rb} \nn \\
&\times\prod_{n=1}^p \frac{\lb p+r-2n\rb \lb p+r-2(n-1)\rb} {\lb
p+r-2(n-1)-2\la\rb \lb p+r-2n+2\la\rb} \frac{\lb p-r-2(n-1)+2\la\rb
\lb p-r-2n+2\la\rb}{\lb p-r-2(n-1)\rb \lb p-r-2n-2\la\rb} \nn \\
&\times f(\mu-\frac{p+r}2) \label{fprmu}
\eea
\hspace{6cm} (for $p,r = 0,1,2,...\leq \la$).

\vspace{.3cm}
Before we discuss the analytic structure of these S-matrix elements
in the following section, we should mention one interesting result
which emerges here: in accordance with the results
of SG theory (see
\cite{zamol1}) and Smirnov's work on $a_2^{(2)}$ ATFT (see
\cite{smirn2}) it seems natural to
identify the two lowest breather states ($B_1$ and $\ol B_1$) with the
two fundamental quantum particles, which are the only states present
in real ATFT. The two fundamental particles of real $a_2^{(1)}$ ATFT
are mass degenerate and are
conjugate to each other.

Hollowood calculated the first quantum mass corrections for the
solitons of $a_n^{(1)}$-ATFT in \cite{hollo3}. Using the conjectured
soliton S-matrix he was able to show that the masses of the quantum
particles (calculated up to one-loop order) are indeed identical to
the masses of the lowest breather states including these first quantum
corrections.
We are now in a position to further justify this identification
by comparing the breather-breather S-matrix elements with the real ATFT
S-matrix given by Braden et al.\ in \cite{brade1}.
If we consider the two lowest breather states $B_1$ and $\ol B_1$, we
get from formula (\ref{brbr}):
\bea
S_{B_1B_1}(\th) &=&
\Bigl(2\Bigr) \Bigl(2\la\Bigr) \Bigl(-2-2\la\Bigr) \nn \\
S_{\ol B_1B_1}(\th) &=&
\Bigl(-2+3\la\Bigr) \Bigl(3\la\Bigr) \Bigl(2+5\la\Bigr) \Bigl(\la
\Bigr). \nn
\eea
If we change the imaginary coupling constant from $i\beta$
to $\beta$ ($\beta$ real) in these two functions, then we indeed
obtain the expression of the conjectured exact S-matrix for real
\mbox{$a_2^{(1)}$ ATFT} (see \cite{brade1}).
Thus the lowest breather states $B_1$ and $\ol B_1$ in the complex
theory can indeed be identified with the two fundamental quantum
particles 1 and 2. This fact also gives further support to the
quantum group S-matrix conjecture by Hollowood.
However, a full understanding of the origin and the implications of
this duality between quantum particles and bound states of solitons is
still missing, and further work is needed.
\section{Pole Structure}

In this section we attempt to explain the poles in the
S-matrix elements obtained in the previous section. (Here we shall
take $\la$ to be sufficiently large to admit several breathers and
excited
solitons into the spectrum. We also assume that $3\la$ is not an
integer, since in this case different poles frequently coincide and it
becomes rather more complicated to explain the entire pole structure.
The case in which $3\la$ or $\la$ are integers requires separate
examination.)

We will conjecture a complete set of three-point couplings of the
theory and try to explain most
other poles either by simple tree level and box diagrams or by a
generalized Coleman-Thun mechanism.
Which poles correspond to fusion processes can be
determined by using the bootstrap principle.
The fusions of two elementary solitons into bound states are
illustrated
in {\em Figure 2a} and {\em 2b}. Using the mass formula
(\ref{mass}) we can calculate the masses of breathers and excited
solitons from their corresponding S-matrix pole and obtain (see
\cite{hollo2}):
\bea
m_{B_p} &=& 2m_A\sin(\frac{\pi p}{3\la}) \nn \\
m_{A_p} &=& 2m_A\cos(\frac{\pi}3 - \frac{\pi p}{3\la}). \label{masses}
\eea
In the same way formula (\ref{mass}) can be used to find other three
point couplings by looking at the poles in higher S-matrix elements.
So for instance, the process in {\em Fig.2d} can exists, since the
element $S^I_{r,p}(i\pi-\th)$ contains a pole $\th =
-\frac{i\pi}{3\la}(p-r)+i\pi$ and it is easily checked that:
\[
m_{B_{p-r}}^2 = m_{A_p}^2 + m_{A_r}^2 +
2m_{A_p}m_{A_r}cos(-\frac{\pi}{3\la}(p-r) + \pi).
\]
The existence of the process {\em Fig.2d} can then additionally be
checked by applying the bootstrap equations (\ref{bootstrap}).
Using the same method all other three-point vertices in {\em Figure 2}
are found and the vertices {\em Fig.2a - 2f} are conjectured to be a
complete list of all possible three point vertices in the theory, that
is, all other possible fusion processes are simply crossed or
conjugated versions of
these six vertices. These three-point couplings
will be used below to explain other poles through higher order
on-shell diagrams. (In these and all following diagrams time is
meant to be moving `upwards'.)

\vspace{1cm}
%
%
%%%%  Three Point Vertices  %%%%
%
\begin{center}
\begin{picture}(420,120)(-10,-30)
%
%%%  A-A -> Bp %%%
%
\put(0,0){\line(1,1){40}}
\put(40,40){\line(1,-1){40}}
\put(40,40){\line(0,1){64}}
\put(40,40){\circle{16}}
\put(40,17){\vector(0,1){14}}
\put(22,58){\vector(1,-1){10}}
\put(58,58){\vector(-1,-1){10}}
\put(16,7){\shs{\scs{$-\frac{i 2\pi}{3\la}p+i\pi$}}}
\put(-7,64){\shs{\scs{$\pol p+i\frac{\pi}2$}}}
\put(52,64){\shs{\scs{$\pol p+i\frac{\pi}2$}}}
\put(-5,-10){\shs{\fns{$\ol A^j$}}}
\put(76,-10){\shs{\fns{$A^j$}}}
\put(35,108){\shs{\fns{$B_p$}}}
%
%
%%% A-A -> Ap %%%
%
\put(160,0){\line(1,1){40}}
\put(200,40){\line(1,-1){40}}
\put(200,40){\line(0,1){64}}
\put(200,40){\circle{16}}
\put(200,17){\vector(0,1){14}}
\put(182,58){\vector(1,-1){10}}
\put(218,58){\vector(-1,-1){10}}
\put(175,7){\shs{\scs{$-\frac{i 2\pi}{3\la}p+i\frac{2\pi}3$}}}
\put(153,64){\shs{\scs{$\pol p+i\frac{2\pi}3$}}}
\put(212,64){\shs{\scs{$\pol p+i\frac{2\pi}3$}}}
\put(155,-10){\shs{\fns{$A^k$}}}
\put(236,-10){\shs{\fns{$A^l$}}}
\put(195,108){\shs{\fns{$\ol A_p^j$}}}
%
%%%  Bp-Br -> Bp+r %%%
%
\put(320,0){\line(1,1){40}}
\put(360,40){\line(1,-1){40}}
\put(360,40){\line(0,1){64}}
\put(360,40){\circle{16}}
\put(360,17){\vector(0,1){14}}
\put(342,58){\vector(1,-1){10}}
\put(378,58){\vector(-1,-1){10}}
\put(342,7){\shs{\scs{$\pol (p+r)$}}}
\put(313,64){\shs{\scs{$-\pol p+i\pi$}}}
\put(372,64){\shs{\scs{$-\pol r+i\pi$}}}
\put(315,-10){\shs{\fns{$B_r$}}}
\put(396,-10){\shs{\fns{$B_p$}}}
\put(355,108){\shs{\fns{$B_{r+p}$}}}
\put(17,-30){\shs {\em Figure 2a}}
\put(177,-30){\shs {\em Figure 2b}}
\put(337,-30){\shs {\em Figure 2c}}
\end{picture}

\vspace{1.5cm}
\begin{picture}(420,120)(-10,-30)
%
%%%  Ap-Ar -> Bp-r %%%
%
\put(0,0){\line(1,1){40}}
\put(40,40){\line(1,-1){40}}
\put(40,40){\line(0,1){64}}
\put(40,40){\circle{16}}
\put(40,17){\vector(0,1){14}}
\put(22,58){\vector(1,-1){10}}
\put(58,58){\vector(-1,-1){10}}
\put(11,5){\shs{\scs{$-\pol (p-r)+i\pi$}}}
\put(-10,64){\shs{\scs{$-\pol r+i\frac{5\pi}6$}}}
\put(52,64){\shs{\scs{$\pol p+i\frac{\pi}6$}}}
\put(-5,-10){\shs{\fns{$A_p^j$}}}
\put(76,-10){\shs{\fns{$\ol A_r^j$}}}
\put(35,108){\shs{\fns{$B_{p-r}$}}}
%
%
%%% Ap-Ar -> Br-p %%%
%
\put(160,0){\line(1,1){40}}
\put(200,40){\line(1,-1){40}}
\put(200,40){\line(0,1){64}}
\put(200,40){\circle{16}}
\put(200,17){\vector(0,1){14}}
\put(182,58){\vector(1,-1){10}}
\put(218,58){\vector(-1,-1){10}}
\put(170,5){\shs{\scs{$-\pol (r-p)+i\pi$}}}
\put(153,64){\shs{\scs{$\pol r+i\frac{\pi}6$}}}
\put(209,64){\shs{\scs{$-\pol p+i\frac{5\pi}6$}}}
\put(155,-10){\shs{\fns{$A_p^j$}}}
\put(236,-10){\shs{\fns{$\ol A_r^j$}}}
\put(195,108){\shs{\fns{$B_{r-p}$}}}
%
%%%  Ap-Ap -> Ap %%%
%
\put(320,0){\line(1,1){40}}
\put(360,40){\line(1,-1){40}}
\put(360,40){\line(0,1){64}}
\put(360,40){\circle{16}}
\put(360,17){\vector(0,1){14}}
\put(342,58){\vector(1,-1){10}}
\put(378,58){\vector(-1,-1){10}}
\put(355,7){\shs{\scs{$i\frac{2\pi}3$}}}
\put(328,64){\shs{\scs{$i\frac{2\pi}3$}}}
\put(377,64){\shs{\scs{$i\frac{2\pi}3$}}}
\put(315,-10){\shs{\fns{$A_p^k$}}}
\put(396,-10){\shs{\fns{$A_p^l$}}}
\put(355,108){\shs{\fns{$\ol A_p^j$}}}
\put(17,-30){\shs {\em Figure 2d}}
\put(177,-30){\shs {\em Figure 2e}}
\put(337,-30){\shs {\em Figure 2f}}
\end{picture}
\end{center}

\vspace{0.3cm}
The (imaginary) angles in the diagrams represent the rapidity
difference of the particles. The sum of the values of the rapidities
in each diagram is $2\pi i$.

\vspace{.3cm}
As mentioned above, all poles in the S-matrix elements describing the
scattering of elementary solitons were already explained by fusion
into bound states. In order to explain the mechanisms responsible for
the occurence of poles in other S-matrix elements, we will
discuss the pole structure of the elements $S_{B_pA}$ and $S_{AB_p}$
in detail. For poles in all other S-matrix elements we will restrict
ourself to giving the relevant diagrams in Appendix B. \vspace{0.3cm}
\\
{\bf \ul{ i) $S_{B_pA}$}}\\
$S_{B_pA}$ contains the following poles:
\bea
\th_n &= -\pol (p-2n) + i\frac{\pi}2 \hsv \hsv &(\mbox{simple for
}n=0,p) \nn \\
&&(\mbox{double for } n=1,2,...,p-1) \label{SBpApoles}
\eea
The simple pole $\th_p$ corresponds to the fusion process $B_p + A^j
\to A^j$, and is therefore explained by the tree-level diagram {\em
Fig.3a}. Since $\th_o = i\pi - \th_p$, the other simple pole is then
simply explained by the crossed version of this tree-level process.

To explain double poles in a two dimensional S-matrix one has to
consider box diagrams like {\em Fig.3b} or crossed box diagrams like
{\em Fig.4}. The Cutkosky rules in two dimensions state that in an
on-shell diagram every loop contributes a two-dimensional integral
whereas
every internal line contributes a delta-function \cite{eden}. Thus, if
diagrams like {\em Fig.3b} or {\em Fig.4} with all internal particles
on-shell exist, then they correspond to double poles in the direct
channel of the S-matrix. This phenomenon was first observed by Coleman
and Thun in \cite{colem} for the Sine-Gordon theory, and is therefore
called the Coleman-Thun mechanism.

Since the (purely imaginary) rapidity differences correspond to (real)
angles in the diagrams, we can use elementary Euclidean
geometry to study higher order processes, such as
{\em Fig.3b}, {\em Fig.4} or any of the diagrams in Appendix D.
All angles in diagram {\em Fig.3b} are fixed by the
three point vertices ({\em Fig.2a-f}). It is therefore
straightforward to calculate the rapidity difference of the
incoming particles. We obtain that the process
pictured in {\em Fig.3b} can occur if the incoming particles $B_p$ and
$A^j$ have a rapidity difference of \mbox{$\th =
-\pol (p-2s) + i\frac{\pi}2$}. Since $s$ can take values $1\leq s \leq
p-1$, these values of the rapidity difference are exactly the double
poles in (\ref{SBpApoles}). Thus all poles in $S_{B_pA}$ are
explained by the two scattering processes in {\em Fig.3}.

\vspace{1.3cm}
\begin{center}
\begin{picture}(330,160)(-10,-30)
%
%%%%%%%% Bp-A Tree %%%%%%%%%%%%%%%%%%%%%%%
\put(0,0){\line(1,1){40}}
\put(80,0){\line(-1,1){40}}
\put(0,140){\line(1,-1){40}}
\put(80,140){\line(-1,-1){40}}
\put(40,40){\line(0,1){60}}
\put(-5,-10){\shs{\fns{$B_p$}}}
\put(75,-10){\shs{\fns{$A^j$}}}
\put(-5,142){\shs{\fns{$A^j$}}}
\put(75,142){\shs{\fns{$B_p$}}}
\put(25,68){\shs{\fns{$A^j$}}}
%
%%%%%%%%% Bp-A  Box %%%%%%%%%%%%%%%%%%%%%%%
%
\put(200,0){\line(1,2){15}}
\put(300,0){\line(-1,2){25}}
\put(215,30){\line(3,1){60}}
\put(215,30){\line(1,6){10}}
\put(275,50){\line(1,6){10}}
\put(225,90){\line(3,1){60}}
\put(200,140){\line(1,-2){25}}
\put(300,140){\line(-1,-2){15}}
\put(195,-10){\shs{\fns{$B_p$}}}
\put(295,-10){\shs{\fns{$A^j$}}}
\put(295,142){\shs{\fns{$B_p$}}}
\put(195,142){\shs{\fns{$A^j$}}}
\put(197,60){\shs{\fns{$B_{p-s}$}}}
\put(284,76){\shs{\fns{$A^j$}}}
\put(243,31){\shs{\fns{$B_s$}}}
\put(251,104){\shs{\fns{$\ol A^j$}}}
\put(20,-30){\shs{\em Figure 3a}}
\put(225,-30){\shs{\em Figure 3b}}
\end{picture}
\end{center}
\newpage
\hspace{-0.9cm} {\bf \ul{ ii) $S_{AB_p}$}}\\
$S_{AB_p}$ contains no double poles in general, but the following
simple poles:
\bea
\th_n^{(a)} &=& -\pol (p-2n) + i\frac{\pi}6 \nn \\
\th_n^{(b)} &=& -\pol (p-2(n-1)) +i\frac{5\pi}6 \hsv \hsv (\mbox{for}
\hfv n=1,2,...,p)
\label{SABppoles}
\eea
(it should be noted that for certain values of $n$ some of these
poles might not lie on the physical strip, and can therefore be ignored).

The simple poles $\th_p^{(a)}$ and $\th_1^{(b)}$ correspond to the
fusion process $A^j + B_p \to A^j_p$ (see {\em Fig.2d} for $r=0$) in
the direct and crossed channel respectively.
In order to explain all other poles we encounter a new phenomenon, the
so-called generalized
Coleman-Thun mechanism, explained in detail in
\cite{corri1}. This mechanism corresponds to diagrams like {\em
Fig.4}, which initially would be expected to lead to double poles, as
explained above. However, we need to consider the transition process
of the two particles in the center of the diagram (indicated by a
black circle in the diagram). If it happens that the
S-matrix element of this process exhibits a simple zero at exactly the
right
rapidity difference, then this zero reduces the
expected double pole, and a diagram like {\em Fig.4} leads to a simple
pole in the direct channel of $A^j-B_p$ scattering. This mechanism
occurs in the case of $S_{AB_p}$.

\vspace{1cm}
%
%%%%  A-Bp crossed box  %%%%
%
\begin{center}
\begin{picture}(140,160)(-10,-30)
\put(0,0){\line(1,2){20}}
\put(120,0){\line(-1,2){20}}
\put(20,40){\line(0,1){60}}
\put(20,40){\line(4,3){80}}
\put(20,100){\line(4,-3){80}}
\put(100,40){\line(0,1){60}}
\put(20,100){\line(-1,2){20}}
\put(100,100){\line(1,2){20}}
\put(60,70){\circle*{8}}
\put(-5,-10){\shs{\fns{$A^j$}}}
\put(115,-10){\shs{\fns{$B_p$}}}
\put(115,142){\shs{\fns{$A^j$}}}
\put(-5,142){\shs{\fns{$B_p$}}}
\put(7,65){\shs{\fns{$B_s$}}}
\put(102,65){\shs{\fns{$B_s$}}}
\put(40,46){\shs{\fns{$A^j_s$}}}
\put(68,46){\shs{\fns{$B_{p-s}$}}}
\put(36,89){\shs{\fns{$B_{p-s}$}}}
\put(71,89){\shs{\fns{$A^j_s$}}}
\put(30,-30){\shs{\em Figure 4}}
\end{picture}
\end{center}

\vspace{0.3cm}
Using diagrams {\em Fig.2a-f} and elementary geometrical calculations
we can compute all angles in diagram {\em Fig.4} and obtain
for the rapidity difference of the incoming particles $A^j$ and
$B_p$:
\[
\th = -\pol (2s-p) +i\frac{5\pi}6 \hspace{1cm} (=\th_{p+1-s}^{(b)})
\]
in which s can take values $1\leq s \leq p-1$.
The rapidity difference for the incoming internal particles $A^j_s$
and $B_{p-s}$ is $\th
= \pol p + i\frac{5\pi}6$, which is a simple zero in
$S_{A_sB_{p-s}}(\th)$. Thus it emerges
that the transition process in the center of this diagram indeed
occurs at a rapidity difference where the corresponding transition
amplitude
exhibits a simple zero, which reduces the expected double poles to
simple poles. Thus we reach the result that all remaining simple poles
$\th_n^{(b)}$ ($n=2,3,...,p$) in
$S_{AB_p}$ are explained by this generalized Coleman-Thun mechanism.
Since $S_{AB_p}$ is itself crossing symmetric, the poles $\th_n^{(a)}$
are just cross channel poles of $\th_{p-n}^{(b)}$ and therefore
correspond to the same process in the crossed channel.

In the following we will restrict ourself to giving the poles of the
S-matrix elements and refer to the corresponding diagrams in Appendix
D. \vspace{0.3cm} \\
{\bf \ul{ (iii) $S_{B_rB_p}$}}\\
For the sake of simplicity let us assume here $r \geq p$.
$S_{B_rB_p}$ contains the poles:
\bea
\th^{(a)}_n =& \pol (p+r-2n) \hsv &(\mbox{simple for } n = 0,p) \nn \\
&&(\mbox{double for } n = 1,2,...,p-1) \nn \\
\th^{(b)}_n =& -\pol(r-p+2n)+i\frac{4\pi}3 &(\mbox{simple for } n =
1,2,...,p) \nn \\
\th^{(c)}_n =& -\pol(r-p+2(n-1))+i\frac{2\pi}3 &(\mbox{simple for } n
= 1,2,...,p).
\eea
The two simple poles $\th^{(a)}_0$ and $\th^{(a)}_p$ are due to the
fusion processes $B_r + B_p \to B_{r+p}$ in the direct channel and
$B_p + \ol B_r \to \ol B_{r-p}$ in the cross channel ({\em Fig.2c}),
whereas the double poles $\th^{(a)}_n$ are explained by the crossed
box diagram in Appendix D ({\em Fig.5a}). Unlike in the former case,
here the transition amplitude does not have a zero, so that this
diagram leads to the expected double pole in the S-matrix.

There are neither tree level nor box diagrams which could explain the
remaining poles in $S_{B_rB_p}$, but there exist higher order
diagrams. We will give one example of a third order process.
In the diagram, depicted in Appendix D {\em Fig.5b}, the rapidity
difference of the incoming particles is equal to $\th_n^{(b)}$ (for
$s=p-n$). This diagram contains
$5$ loops and $13$ internal lines, and should therefore lead to a
triple pole. However, a careful study of all internal angles reveals
that two of the three transition processes in the diagram occur at a
rapidity difference, where the corresponding transition amplitudes
exhibit a simple zero. These zeros reduce the expected triple poles to
the simple
poles $\th^{(b)}_n$ ($n=1,2,...p$). We were not able to find a second
or third order diagram which could explain the remaining string of
poles $\th^{(c)}_n$. We expect these poles to correspond to
fourth or even higher order diagrams.
\vspace{0.3cm} \\
{\bf \ul{ (iv) $S_{A_rB_p}$}} \\
This element contains four simple poles and two strings of double
poles:
\bea
\th^{(a)}_n =& -\pol(r+p-2n) + i\frac{5\pi}6 \hsv &(\mbox{simple for }
n = 0,p) \nn \\
&&(\mbox{double for } n = 1,2,...,p-1) \nn \\
\th^{(b)}_n =& -\pol(p-r-2n)+i\frac{\pi}6 &(\mbox{simple for }
n = 0,p) \nn \\
&&(\mbox{double for } n = 1,2,...,p-1)
\eea

The four simple poles can be explained by tree level processes.
$\th^{(b)}_p$ and $\th^{(a)}_0$ correspond to the fusion process
$A^j_r + B_p \to A^j_{r+p}$ ({\em Fig.2d}) in the direct and crossed
channel. $\th^{(a)}_p$ and $\th^{(b)}_0$ correspond to the fusion
process $A^j_r + B_p \to A^j_{r-p}$ ({\em Fig.2e}).
The double poles are due to the process illustrated by the box diagram
{\em Fig.6} in the direct and crossed channel.
\vspace{0.3cm} \\
{\bf \ul{ (iv) $S_{B_pA_r}$}}\\
This element contains four strings of simple poles:
\bea
\th^{(a)}_n &=& -\pol(r+p-2(n-1)) +i\frac{7\pi}6 \nn \\
\th^{(b)}_n &=& -\pol(p-r-2n) -i\frac{\pi}6 \nn \\
\th^{(c)}_n &=& -\pol(r+p-2n)+i\frac{\pi}2 \nn \\
\th^{(d)}_n &=& -\pol(p-r-2(n-1))+i\frac{\pi}2 \hsv \hsv (\mbox{for }
n = 1,2,...,p). \nn
\eea
Since $\th^{(a)}$ and $\th^{(b)}$, as well as $\th^{(c)}$ and
$\th^{(d)}$, are crossed channel poles to each other it is sufficient
to find corresponding diagrams to only two of these strings of poles.
$\th^{(d)}_1$ (and therefore also $\th^{(c)}_p$) can be explained by
the process in {\em Fig.7a}. The diagram {\em Fig.7b} explains the
poles $\th^{(a)}_n$ and $\th^{(b)}_n$ ($n=1,2,...,p$), whereas the
remaining poles
$\th^{(c)}_n$ and $\th^{(d)}_{p-n}$ (for $n=2,3,...,p$) are due to
the third order process in {\em Fig.7c}, which displays (similar to
{\em Fig.5b}) a `double' generalized Coleman-Thun mechanism.
\vspace{0.3cm} \\
{\bf \ul{ (iv) $S_{A_pA_r}$}}\\
The S-matrix elements for the scattering of excited solitons exhibit
a large number of simple and multiple poles. We were not able to
explain all poles in a general way as done for the other S-matrix
elements. However, we were able to explain the pole structure for some
simple explicit examples through low order diagrams and the
generalized Coleman-Thun mechanism by using the above list of three
point couplings.
We expect that all simple and higher order poles in these
S-matrix elements can be explained in this way. However, in many
cases the relevant diagrams can become extremely complicated and may
therefore prove more elusive. \vspace{0.3cm}

We do not have a direct way of proving that the list of three-point
couplings in \mbox{{\em Figure 2a-f}}, is complete and that there are
no other fusion processes possible. However, the fact that we were
able to explain almost all other poles in the S-matrix through higher
order diagrams using only those six three-point vertices provides
strong evidence for our conjecture that the list of three point
couplings is indeed complete.

\section{Discussion}

Apart from SG-theory ($a_1^{(1)}$ ATFT), the complete spectra of
solitons and bound
states in any other complex ATFT have been heretofore unknown. In this
paper we have attempted to solve this problem for the subsequent (and
more complicated) $a_2^{(1)}$ affine Toda field theory.
We have constructed bound states of elementary solitons and
conjectured that these bound states together with the solitons and
their antisolitons form the complete spectrum of the
theory. We have calculated explicit S-matrix elements for the
scattering of solitons and bound states and have been able to explain
their pole structure by using a surprisingly small number of three
point couplings. The S-matrix elements for the lowest breather states
have been shown to coincide with the S-matrix for the fundamental
quantum particles, which provides a further justification for this
S-matrix conjecture.

The discussion in this paper has been restricted to the case where
$3\la$ is not an integer. This case should be considered separately.

It is necessary to further consider the question of unitarity.
Although the theory is non-unitary in general
Hollowood suggested in \cite{hollo2} that the S-matrix
seems to describe a unitary field theory in the strong coupling limit
(i.e. if $\la$ is sufficiently small). One could also ask whether certain
restrictions of the theory, like RSOS restrictions or the restriction
to integer values of $\la$, would lead to unitary theories.

Another question unmentioned in this paper relates to the nature of
the possible connection of the
residues of poles and the parity of bound states with the problem of
unitarity, first discussed in \cite{karow}. The task of
calculating the pole residues of our S-matrix elements seems rather
difficult, but it would be interesting to examine in what way the pole
residues change sign with the change of $\la$ (an interesting
connection of the change of residue signs with the generalized
Coleman-Thun mechanism has been discovered in \cite{corri1}).

The question arises as to whether our method is
equally applicable to the more general case of $a_n^{(1)}$ ATFT's.
Although more elaborate, this application seems to be possible in
principle.
However, an additional problem surfaces for $n \geq 3$,
namely the problem of unfilled weights of the fundamental
representations. This problem has been examined in detail by McGhee in
\cite{mcghe} and it emerged that for the $a_n^{(1)}$ series of ATFT
only in the case $n=1$ and $n=2$ the topological charges of the
soliton solutions fill
out the weight lattice of the fundamental representations. It is
presently not clear how in the quantum case this could affect the
conjectured S-matrix.

Another possible area for further research would be the search for a
more rigorous
definition of the generators and commutation relations of the
Zamolodchikov algebra (\ref{commrel}), as was done in terms of
vertex operators in \cite{corri2}.

The question also remains as to how to construct a consistent
soliton S-matrix for theories based on other algebras and in
particular on non-simply laced or twisted algebras.
In two recent publications (\cite{macka2}, \cite{deliu2}) the first
quantum mass corrections of affine Toda solitons were calculated by
using semiclassical techniques. It was found that for many theories
the soliton masses do not renormalise according to the masses of the
quantum particles. In these cases the approach of defining an S-matrix
in terms of the basic R-matrix, seems not to be applicable in the same
way as for the $a_n^{(1)}$ theories. A possible solution to this
problem may lie in using R-matrices with a different gradation (see
\cite{babic}).
\vspace{1cm} \\
{\large {\bf Acknowledgement}}\\
I would like to thank N.J. MacKay for introducing me to this subject
and for many discussions during the course of this work.
%
%
%
%%%%%%%%%%%%%%%%%%%%%%%%%%%%%%%%%%%%%%%%%%%%%%%%%%%%%%%%%%%%%%%%
\newpage
\begin{flushleft}
{\large {\bf APPENDIX A: Sine-Gordon S-Matrix}}
\vspace{.3cm} \\
\end{flushleft}
In this appendix the complete S-matrix for the Sine-Gordon model, as
derived in \cite{zamol1}, is given. We use the following notations,
which helps to illustrate the connection with our S-matrix elements
for $a_2^{(1)}$-ATFT.
\bea
\lb x \rb_S &\equiv& sin(\frac{\th}{2i} + \frac{\pi x}{4\la}) \nn \\
\bigl( x \bigr)_S &\equiv& \frac{\lb x \rb_S}{\lb -x \rb_S}. \nn
\eea
For convenience we also use the abbreviations $\tilde{\mu} \equiv
i\frac{\la}{\pi}\th$ and $\la \equiv \frac{4\pi}{\beta^2}-1$.
\\
The commutation relations for the complete Zamolodchikov algebra are
the following:
\bea
A(\th_1)A(\th_2) &=& \tilde{S}^I(\th_{12}) A(\th_2)A(\th_1)
\nn \\
A(\th_1)\ol A(\th_2) &=& \tilde{S}^T(\th_{12}) \ol A(\th_2)A(\th_1) +
\tilde{S}^R(\th_{12}) A(\th_2)\ol A(\th_1) \nn \\
A(\th_1)B_p(\th_2) &=& \tilde{S}_p(\th_{12})B_p(\th_2)A(\th_1) \nn \\
B_r(\th_1)B_p(\th_2) &=& \tilde{S}_{r,p}(\th_{12})B_p(\th_2)B_r(\th_1)
\nn
\eea
and the S-matrix elements are given by:
\bea
\tilde{S}^I(\th) & = & 2i \sin(i\la\th + \la\pi)f_{SG}(\tilde{\mu})
\nn \\
\tilde{S}^T(\th) & = & 2i \sin(-i\la\th) f_{SG}(\tilde{\mu}) \nn \\
\tilde{S}^R(\th) & = & 2i \sin(\la\pi)f_{SG}(\tilde{\mu}) \nn \\
\tilde{S}_p(\th) &=& \prod_{n=1}^p\frac{\lb 2n-p-3\la\rb_S \lb
p-2n+\la\rb_S} {\lb p-2n-\la\rb_S \lb 2n-p-\la\rb_S} \nn \\
\tilde{S}_{r,p}(\th) &=&
\prod_{n=1}^p\biggl(r+p-2n\biggr)_S\biggl(r+p-2(n-1)\biggr)_S
\biggl(p-r-2n-2\la\biggr)_S\biggl(p-r-2(n-1)+2\la\biggr)_S. \nn
\eea \vspace{0.3cm} \\
The scalar factor $f_{SG}(\tilde{\mu})$ ensures crossing symmetry and
can be written in the following form:
\bea
f_{SG}(\tilde{\mu}) = \frac1{2i \sin(\pi(\tilde{\mu}+\lambda))}
\prod_{j=1}^{\infty} \frac{\Gamma[1+\tilde{\mu}+(2j-2)\lambda]
\Gamma[\tilde{\mu}+2j\lambda]} {\Gamma[1-\tilde{\mu}+(2j-2)\lambda]
\Gamma[-\tilde{\mu}+2j\lambda]} \nn \\
\times\frac{\Gamma[-\tilde{\mu}+(2j-1)\lambda]
\Gamma[1-\tilde{\mu}+(2j-1)\lambda]}
{\Gamma[\tilde{\mu}+(2j-1)\lambda]
\Gamma[1+\tilde{\mu}+(2j-1)\lambda]}. \nn
\eea
%
%
%%%%%%%%%%%%%%%%%%%%%%%%%%%%%%%%%%%%%%%%%%%%%%%%%%%%%%%%%%%%%%%%%%%%%%
%
\newpage
\begin{flushleft}
{\large {\bf APPENDIX B: Calculation of $S_{AB_p}$}}
\vspace{.3cm} \\
\end{flushleft}
In this appendix we show the calculations used to obtain the
commutation relation of $A^j(\th)$ ($j=1,2,3$) with $B_p(\th)$.  For
convenience let us define the breather pole:
\[
\Th_p \equiv -i\frac{2\pi}{3\la}p+i\pi
\]

Regarding the formal definition (\ref{brdef}) of the symbols $B_p$, we
need to consider the triple product $A^j(\th_1)
\sum_{m=1}^3[\al_m^{(p)} A^m(\th_2)\ol A^m(\th_3)]$.
Using the commutation relations for elementary solitons
and their antisolitons (\ref{commrel}) and (\ref{ancommrel}), we obtain
the following expression.
\bea
\lefteqn{A^j(\th_1)\sum_{m=1}^3[\al_m^{(p)}A^m(\th_2)\ol A^m(\th_3)]
=} \nn \\ & &\al_j^{(p)}\biggl\{S^I(\th_{12})S^I(i\pi-\th_{13}) +
\frac{\al_k^{(p)}}{\al_j^{(p)}}
S^{R(j,k)}(\th_{12})S^{R(j,k)}(i\pi-\th_{13}) \nn \\ & &
\hspace{2cm} + \frac{\al_l^{(p)}}{\al_j^{(p)}}
S^{R(j,l)}(\th_{12})S^{R(j,l)}(i\pi-\th_{13})\biggr\}_{(1)}
A^j(\th_2)\ol A^j(\th_3) A^j(\th_1) \nn \\ & & + \al_k^{(p)}
\biggl\{S^T(\th_{12})S^T(i\pi-\th_{13})\biggr\}_{(2)} A^k(\th_2)\ol
A^k(\th_3) A^j(\th_1) \nn \\ & & + \al_l^{(p)}
\biggl\{S^T(\th_{12})S^T(i\pi-\th_{13})\biggr\}_{(3)} A^l(\th_2)\ol
A^l(\th_3) A^j(\th_1) \nn \\ & & + \biggl\{\al_j^{(p)}
S^I(\th_{12})S^{R(k,j)}(i\pi-\th_{13}) +
\al_k^{(p)} S^{R(j,k)}(\th_{12})S^I(i\pi-\th_{13}) \nn \\
& & \hspace{2cm} + \al_l^{(p)} S^{R(j,l)}(\th_{12})
S^{R(k,l)}(i\pi-\th_{13})\biggr\}_{(4)} A^j(\th_2)\ol A^k(\th_3)
A^k(\th_1) \nn \\ & & + \biggl\{\al_j^{(p)}
S^I(\th_{12})S^{R(l,j)}(i\pi-\th_{13}) +
\al_l^{(p)} S^{R(j,l)}(\th_{12})S^I(i\pi-\th_{13}) \nn \\
& & \hspace{2cm} + \al_k^{(p)} S^{R(j,k)}(\th_{12})
S^{R(l,k)}(i\pi-\th_{13})\biggr\}_{(5)} A^j(\th_2)\ol
A^l(\th_3)A^l(\th_1).  \label{app1}
\eea
In order to obtain the required scattering amplitude we have to take the
limit ($\th_3-\th_2 \to \Th_p$) on both sides of this expression.
By taking this limit we obtain the following and rather simple
identities for the factors on the right hand side of (\ref{app1}):
\bea
\lim_{\th_3-\th_2 \to \Th_p}\biggl\{ \hfv \biggr\}_{(1)} &=&
\lim_{\th_3-\th_2 \to \Th_p} \biggl\{ \hfv \biggr\}_{(2)} =
\lim_{\th_3-\th_2 \to \Th_p} \biggl\{ \hfv \biggr\}_{(3)}; \nn \\
\mbox{and} \hspace{1cm}
\lim_{\th_3-\th_2 \to \Th_p} \biggl\{ \hfv \biggr\}_{(4)} &=&
\lim_{\th_3-\th_2 \to \Th_p} \biggl\{ \hfv \biggr\}_{(5)}
= 0. \nn
\eea
Since the last two terms vanish, we see that the only scattering
process possible is a transition of the two incoming particles.
Hence we introduce the commutation relation:
\[
A^j(\th_1)B_p(\frac{\th_2+\th_3}2) = S_{AB_p}(\th)
B_p(\frac{\th_2+\th_3}2)A^j(\th_1)
\]
in which $\th = \th_1 - \frac{\th_2+\th_3}2$.
According to expression (\ref{app1}) the scalar function
$S_{AB_p}(\th)$ must be
\[
S_{AB_p}(\th) \equiv \lim\biggl\{ \biggr\}_{(2)} = \lim_{\th_3-\th_2 \to
\Th_p} S^T(\th_{12}) S^T(i\pi-\th_{13})
\]
or in terms of $\th$:
\bea
S_{AB_p}(\th) &=& S^T(\th+\frac12\Th_p) S^T(i\pi-\th+\frac12\Th_p)
\nn \\
&=& (-4)\sin(-i\frac32\la\th - \frac{\pi p}2 + \frac34\la\pi)
\sin(i\frac32\la\th -\frac{\pi p}2 +\frac94\la\th) \nn \\
& & \times f_{0,0}(\mu+\frac p2-\frac34\la) f_{0,0}(-\mu +\frac
p2-\frac94\la). \label{app2}
\eea

The calculation of the other S-matrix elements, listed in section 4,
proceeds on similar lines.
\vspace{1.5cm} \\
{\large {\bf APPENDIX C: Some properties of the scalar factor}}
\vspace{.3cm} \\
In order to simplify expressions such as (\ref{app2}) we need to state
some
properties of the scalar factor $f_{0,0}(\mu)$, which has been
computed by Hollowood in \cite{hollo2} to be $f_{0,0}(\mu) = \bigl(
2\la\bigr) f(\mu)$ and
\bea
f(\mu) = \frac1{2i \sin(\pi(\mu+\lambda))} \prod_{j=1}^{\infty}
\frac{\Gamma[1+\mu+(3j-3)\lambda] \Gamma[\mu+3j\lambda]}
{\Gamma[1-\mu+(3j-3)\lambda] \Gamma[-\mu+3j\lambda]} \nn \\
\times\frac{\Gamma[-\mu+(3j-2)\lambda] \Gamma[1-\mu+(3j-1)\lambda]}
{\Gamma[\mu+(3j-2)\lambda] \Gamma[1+\mu+(3j-1)\lambda]}. \label{fmu}
\eea
\vspace{.2cm} \\
By
using the elementary property of the $\Ga$-function: $\Ga(z+1) =
z\Ga(z)$  (for $z \neq 0,-1,-2,...$), and the product expansion of the
sine-function:
\be
\sin(x) = x \prod^{\infty}_{n=1}(1-\frac{x^2}{k^2\pi^2}) \nn
\ee
we can derive the following properties of $f(\mu)$
\bea
f(\mu-p) &=& (-1)^p\prod^p_{n=1} \frac{\lb 2(n-1) +2\la \rb\lb 2n-2\la
\rb} {\lb 2(n-1) \rb\lb 2n \rb} f(\mu) \hspace{10pt} \nn \\
f(\mu)f(-\mu-3\la) &=& \frac1{4\sin(\pi\mu)\sin(\pi(\mu+3\la))} \nn \\
f(\mu)f(\mu-\la) &=& \frac{-\sin(\pi(\mu-3\la))}{2i\sin(\pi(\mu+\la))
\sin(\pi(\mu-2\la))} f(-\mu+2\la) \label{fprop}
\eea
in which $p$ is a positive integer and the notation $[x]$ has been
used as defined in section 4.\\
Additionally we trivially have
\be
f(-\mu) = \frac1{-4\sin^2(\pi(\mu+\la))}(f(\mu))^{-1}.
\ee
It should be mentioned that we assumed in these calculations that
$\la$ is not an integer. This case has to be considered separately,
since there can appear additional singularities in the function
$f(\mu)$, which seem to render the above calculations invalid.

Using these properties we can see that the infinite
product of Gamma-functions in (\ref{app2}) vanishes and we end up with
only a finite product of sine-functions. In the same way the overall
scalar factors of all other S-matrix elements involving the scattering
of breathers can be simplified into finite products, and we obtain the
expressions (\ref{solbr}) - (\ref{exsolbr}) in section~4.

We will briefly show some of the steps in the calculation which lead
to the expression (\ref{fprmu}) of the scalar factor $f_{p,r}(\mu)$.
In order to obtain the commutation relations for the scattering of two
excited solitons, e.g.\ $A^j_p$ and $A^j_r$, we need to consider the
quadruple product:
\be
[\ga_1\ol A^l(\th_1)\ol A^k(\th_2) + \ga_2\ol A^k(\th_1) \ol
A^l(\th_2)] \times [\ga_1\ol A^l(\th_3)\ol A^k(\th_4) + \ga_2\ol
A^k(\th_3) \ol A^l(\th_4)] \label{app3}
\ee
Using the relations (\ref{commrel}), we can work out this expression in
a similar way as done above for $A^j - B_p$ scattering. Taking the
`double' limit: \mbox{$\th_2-\th_1 \to -i\frac{2\pi}{3\la}p +
i\frac{2\pi}3$} and \mbox{$\th_4-\th_3 \to -i\frac{2\pi}{3\la}r +
i\frac{2\pi}3$}, we obtain the overall scalar factor in this product:
\[
f_{0,0}(\mu + \frac{p+r}2-\la) f_{0,0}(\mu + \frac{r-p}2)
f_{0,0}(\mu + \frac{p-r}2) f_{0,0}(\mu - \frac{p+r}2+\la).
\]
By using the above properties (\ref{fprop}) we can show that this
expression is equal to
\[
\frac1{(2i)^3} \frac1{\sin(\pi(\mu+\frac{p+r}2+2\la))
\sin(\pi(\mu+\frac{p+r}2-\la)) \sin(\pi(\mu+\frac{p+r}2))} f_{p,r}(\mu)
\]
in which $\mu = i\frac{3\la}{2\pi}(\frac{\th_1+\th_2}2 -
\frac{\th_3+\th_4}2)$ and $f_{p,r}(\mu)$ has the form (\ref{fprmu})
given in section 4. Considering all factors in the product
(\ref{app3}) we finally obtain the exchange relations
(\ref{exsolexsol}) in section 4.
\newpage
\begin{flushleft}
{\large{\bf APPENDIX D: Figures}}
\end{flushleft}
(Time is moving upwards in all diagrams.)
\vspace{.5cm}
%
%%%%%%%%%%%%%%   Br-Bp crossed box %%%%%%%%%%%%%%%%%%%%%
\begin{center}
\begin{picture}(140,160)(-10,-30)
\put(0,0){\line(1,2){20}}
\put(120,0){\line(-1,2){20}}
\put(20,40){\line(0,1){60}}
\put(20,40){\line(4,3){80}}
\put(20,100){\line(4,-3){80}}
\put(100,40){\line(0,1){60}}
\put(20,100){\line(-1,2){20}}
\put(100,100){\line(1,2){20}}
\put(60,70){\circle*{8}}
\put(-5,-10){\shs{\fns{$B_r$}}}
\put(115,-10){\shs{\fns{$B_p$}}}
\put(115,142){\shs{\fns{$B_r$}}}
\put(-5,142){\shs{\fns{$B_p$}}}
\put(7,65){\shs{\fns{$B_s$}}}
\put(102,65){\shs{\fns{$B_s$}}}
\put(37,46){\shs{\fns{$B_{r-s}$}}}
\put(68,46){\shs{\fns{$B_{p-s}$}}}
\put(36,89){\shs{\fns{$B_{p-s}$}}}
\put(63,89){\shs{\fns{$B_{r-s}$}}}
\put(30,-30){\shs{\em Figure 5a}}
\end{picture}
\end{center}

\vspace{.5cm}
%
%%%%%%%%%%%%% Bp-Br third order %%%%%%%%%%%%%%%
\begin{center}
\begin{picture}(220,170)(-10,-30)
\put(0,0){\line(3,2){24}}
\put(24,16){\line(1,1){128}}
\put(24,16){\line(4,1){88}}
\put(112,37){\line(0,1){86}}
\put(152,16){\line(3,-2){24}}
\put(112,37){\line(2,-1){40}}
\put(0,160){\line(3,-2){24}}
\put(24,144){\line(1,-1){128}}
\put(24,144){\line(4,-1){88}}
\put(112,123){\line(2,1){40}}
\put(152,144){\line(3,2){24}}
\put(88,80){\circle*{8}}
\put(112,56){\circle*{8}}
\put(112,104){\circle*{8}}
\put(-5,-10){\shs{\fns{$\ol B_p$}}}
\put(175,-10){\shs{\fns{$B_r$}}}
\put(-5,163){\shs{\fns{$B_r$}}}
\put(175,163){\shs{\fns{$\ol B_p$}}}
\put(48,48){\shs{\tiny{$\ol A^k$}}}
\put(48,104){\shs{\tiny{$\ol A^j$}}}
\put(70,22){\shs{\tiny{$A^k$}}}
\put(122,22){\shs{\tiny{$A^j$}}}
\put(126,42){\shs{\tiny{$\ol A^j$}}}
\put(115,77){\shs{\tiny{$\ol A^l_s$}}}
\put(130,115){\shs{\tiny{$\ol A^k$}}}
\put(70,134){\shs{\tiny{$A^j$}}}
\put(124,132){\shs{\tiny{$A^k$}}}
\put(60,-30){\shs{\em Figure 5b}}
\end{picture}
\end{center}

\vspace{.5cm}
%
%
%%%%%%%%%%%%%%%%%     Arj- Bp  box   %%%%%%%%%%%%%%%%
\begin{center}
\begin{picture}(130,160)(-10,-30)
\put(0,0){\line(1,2){15}}
\put(100,0){\line(-1,2){25}}
\put(15,30){\line(3,1){60}}
\put(15,30){\line(1,6){10}}
\put(75,50){\line(1,6){10}}
\put(25,90){\line(3,1){60}}
\put(0,140){\line(1,-2){25}}
\put(100,140){\line(-1,-2){15}}
\put(-5,-10){\shs{\fns{$A_r^j$}}}
\put(95,-10){\shs{\fns{$B_p$}}}
\put(95,142){\shs{\fns{$A_r^j$}}}
\put(-5,142){\shs{\fns{$B_p$}}}
\put(6,60){\shs{\fns{$B_s$}}}
\put(82,76){\shs{\fns{$A^j_{p+r-s}$}}}
\put(43,31){\shs{\fns{$A^j_{r-s}$}}}
\put(39,104){\shs{\fns{$\ol B_{p-s}$}}}
\put(30,-30){\shs{\em Figure 6}}
\end{picture}
\end{center}

\newpage
%
%%%%%%%%%%%%%%   Bp-Ar two crossed boxes %%%%%%%%%%%%%%%%%%%%%
\vspace{1cm}
\begin{center}
\begin{picture}(340,160)(-10,-30)
\put(0,0){\line(1,2){20}}
\put(120,0){\line(-1,2){20}}
\put(20,40){\line(0,1){60}}
\put(20,40){\line(4,3){80}}
\put(20,100){\line(4,-3){80}}
\put(100,40){\line(0,1){60}}
\put(20,100){\line(-1,2){20}}
\put(100,100){\line(1,2){20}}
\put(60,70){\circle*{8}}
\put(-5,-10){\shs{\fns{$B_p$}}}
\put(115,-10){\shs{\fns{$A_r^j$}}}
\put(115,142){\shs{\fns{$B_p$}}}
\put(-5,142){\shs{\fns{$A_r^j$}}}
\put(7,65){\shs{\fns{$A_j$}}}
\put(102,65){\shs{\fns{$A_j$}}}
\put(39,46){\shs{\fns{$\ol A^j$}}}
\put(71,46){\shs{\fns{$B_r$}}}
\put(36,89){\shs{\fns{$B_r$}}}
\put(73,89){\shs{\fns{$\ol A^j$}}}
\put(200,0){\line(1,2){20}}
\put(320,0){\line(-1,2){20}}
\put(220,40){\line(0,1){60}}
\put(220,40){\line(4,3){80}}
\put(220,100){\line(4,-3){80}}
\put(300,40){\line(0,1){60}}
\put(220,100){\line(-1,2){20}}
\put(300,100){\line(1,2){20}}
\put(260,70){\circle*{8}}
\put(195,-10){\shs{\fns{$B_p$}}}
\put(315,-10){\shs{\fns{$A_r^j$}}}
\put(315,142){\shs{\fns{$B_p$}}}
\put(195,142){\shs{\fns{$A_r^j$}}}
\put(207,65){\shs{\fns{$\ol B_s$}}}
\put(302,65){\shs{\fns{$\ol B_s$}}}
\put(237,46){\shs{\fns{$B_{p+s}$}}}
\put(268,46){\shs{\fns{$A_{r-s}^j$}}}
\put(236,89){\shs{\fns{$A_{r-s}^j$}}}
\put(263,89){\shs{\fns{$B_{p+s}$}}}
\put(30,-30){\shs{\em Figure 7a}}
\put(230,-30){\shs{\em Figure 7b}}
\end{picture}
\end{center}

\vspace{1cm}
%
%%%%%%%%%%%%% Bp-Ar third order %%%%%%%%%%%%%%%
\begin{center}
\begin{picture}(220,170)(-10,-30)
\put(0,0){\line(3,2){24}}
\put(24,16){\line(1,1){128}}
\put(24,16){\line(4,1){88}}
\put(112,37){\line(0,1){86}}
\put(152,16){\line(3,-2){24}}
\put(112,37){\line(2,-1){40}}
\put(0,160){\line(3,-2){24}}
\put(24,144){\line(1,-1){128}}
\put(24,144){\line(4,-1){88}}
\put(112,123){\line(2,1){40}}
\put(152,144){\line(3,2){24}}
\put(88,80){\circle*{8}}
\put(112,56){\circle*{8}}
\put(112,104){\circle*{8}}
\put(-5,-10){\shs{\fns{$B_p$}}}
\put(175,-10){\shs{\fns{$A_r^j$}}}
\put(-5,163){\shs{\fns{$A_r^j$}}}
\put(175,163){\shs{\fns{$B_p$}}}
\put(38,52){\shs{\tiny{$B_{p-s}$}}}
\put(48,104){\shs{\tiny{$\ol B_r$}}}
\put(70,22){\shs{\tiny{$B_s$}}}
\put(122,20){\shs{\tiny{$A^j$}}}
\put(127,42){\shs{\tiny{$\ol B_r$}}}
\put(113,77){\shs{\tiny{$A^j$}}}
\put(130,116){\shs{\tiny{$B_{p-s}$}}}
\put(70,134){\shs{\tiny{$A^j$}}}
\put(125,136){\shs{\tiny{$B_s$}}}
\put(60,-30){\shs{\em Figure 7c}}
\end{picture}
\end{center}
%
%
%
%%%%%%%%%%%%%%%%%%%%%%%%%%%%%%%%%%%%%%%%%%%%%%%%%%%%%%%%%%%%%%%%%%%%%%
%
\newpage
\end{document}